\documentstyle[12pt,epsfig]{article}
\begin{document}

\centerline{\bf COSMOLOGY WITH A SHOCK-WAVE}

\vspace{.2cm}
\centerline{November 9, 1998}
\vspace{.2cm}
$$\begin{array}{ccc} Joel\ Smoller\footnotemark[1] & Blake\ Temple\footnotemark[2] \nonumber \end{array}$$
\footnotetext[1]{Department of Mathematics, University of Michigan, 
Ann Arbor, MI 48109 smoller@umich.edu;  Supported in part by NSF 
Applied Mathematics Grant No. DMS-980-2370, and in part by the Institute of Theoretical Dynamics (ITD), UC-Davis.}  

\footnotetext[2]{Institute of Theoretical Dynamics (Mathematical Physics) and Department of Mathematics,
UC-Davis, Davis, CA 95616 temple@math.ucdavi.edu;  Supported in part by NSF 
Applied Mathematics Grant No. DMS-980-2473, and in part by the Institute of Theoretical Dynamics (ITD).}

\newtheorem{Theorem}{Theorem}
\newtheorem{Lemma}{Lemma}
\newtheorem{Proposition}{Proposition}
\newtheorem{Corollary}{Corollary}
\large

\begin{abstract}
We construct the simplest solution of the Einstein equations that 
incorporates a shock--wave into a standard Friedmann--Robertson--Walker
metric whose equation of state accounts for the Hubble constant and
the microwave background radiation temperature.  This produces a new 
solution of the Einstein equations from which we are able to derive
estimates for the shock position at present time.  We show that the
distance from the shock--wave to the center of the explosion at present
time is comparable to the Hubble distance.  We are motivated by the
idea that the expansion of the universe as measured by the Hubble
constant might be accounted for by an event more similar to a 
classical explosion than by the well-accepted scenario of the Big
Bang.
\end{abstract}

\renewcommand{\theequation}{\thesection.\arabic{equation}}
\section{Introduction}
\label{Sect1} 
\setcounter{equation}{0}

In the standard model for cosmology it is assumed that, on the largest
scale, the 
{\em entire} universe is expanding at a rate measured by the Hubble
law, \cite{blaugu,long,misnthwh,peeb,wald,wein}.  Hubble's Law correlates recessional velocities 
of galaxies with red-shifts. However, this correlation has only been 
verified for nearby galaxies, 
and it is an extrapolation to apply this law to the entire universe.   
Moreover, it follows from the 
Einstein equations, \cite{hawkel,hawkpe,wald}, that if the
universe is everywhere expanding, then every spacetime point can be traced
back to a singularity in the past, a singularity from which the entire universe 
{\em burst} in an event referred to as the {\em Big Bang}.  The assumption
 that the Hubble Law 
applies to the 
{\em entire} 
universe is referred to as the Cosmological Principle, \cite{peeb}.  
The Cosmolgical
Principle is the starting assumption in the modern theory of cosmology, 
and it is
precisely this principle that forces the singularity
into the standard Big Bang interpretation of the origin of the universe. 
In this paper we explore the possibility that Hubble's Law actually
only measures a {\em localized} expansion of the universe, and not 
the expansion of the entire universe. We demonstrate the consistency
of this possibility by constructing the simplest possible solution of the 
Einstein equations that accounts for the observed Hubble expansion rate
and the correct microwave background radiation temperature, such that
there is a shock--wave present at the leading edge of the expansion.

Our motivation is the idea that
the expansion of the universe, as measured by the Hubble constant,
might be the result of a large scale {\em localized} explosion that 
generated a shock--wave
at the leading edge, not unlike a classical explosion into a static
background, except on an enormously large scale.  If this were true, then
it would place our solar system in a special position relative to
the center of the explosion, and this would violate the so-called 
{\em Copernican Principle}, at least on the scale at which the 
Hubble Law applies.
The Copernican Principle is the statement that the earth is not in a 
\lq\lq special 
place'' in the universe.  This principle justifies 
the standard cosmology based on the Friedmann--
Robertson--Walker (FRW) metric because the FRW metric is the unique metric
that is consistent with the Einstein equations, and is homogeneous and 
isotropic about every point.  The high degree of uniformity of the 
background microwave radiation in all directions, together with the
directional independence of the redshifting of
galaxies, provides the strongest support for the Copernican Principle.   
The idea that there is a shock--wave present at the leading edge of 
that portion of the universe where the Hubble constant applies, 
also violates another basic tenet
of modern cosmology; namely, that we can meaningfully time reverse the
continuum model all the way back to microseconds after the Big Bang.
(For example, the theory of inflation applies to the regime
$10^{-34}$ to $10^{-32}$ seconds after the Big Bang in the 
continuum model, \cite{long}.) 
Indeed, it follows from the
mathematical theory that
shock--waves introduce
a fundamental increase of entropy and
consequent loss of information, \cite{lax,smol}.  Thus, when a shock--wave is 
incorporated
into cosmology, it becomes impossible to reconstruct the 
details of the early explosion from
present data, at least at the level of the continuum model.

The simplest shock--wave model for cosmology is one in which the \lq\lq 
expanding
universe'', inside the shock--wave, is modeled by the standard FRW metric
of cosmology, and the spacetime on the outside is modeled by a
Tolman--Oppenheimer--Volkoff (TOV) metric, (the general relativistic
version of a static fluid
sphere), such that the interface 
in between is an exact, spherically symmetric shock--wave solution of
the Einstein equations, that propagates outward.  The assumption that 
outside the shock--wave is a 
time-independent
spherically symmetric solution is not unreasonable if one imagines 
that the spacetime before
the explosion occured took a long time getting into the pre-explosion 
configuration;  and the assumption that an expanding FRW metric describes the
spacetime behind the shock--wave is consistent with the fact that the
galaxies appear to be uniformly expanding.  In this paper we 
construct such
a model assuming critical expansion, ($k=0$), for the FRW metric, and 
what emerges is a new, 
essentially exact solution
of the Einstein equations.  We
show that reasonable physical requirements on the TOV equation of state
put an interesting constraint on the possible position of the shock--wave 
relative to the center of the explosion. 
Using this constraint, we derive precise estimates for the  
shock position at 
present
time, as predicted by this 
model; that is, at the time in this model at which the Hubble constant and the value of the background
radiation temperature agree with observed values.  
The constraint on the shock position can be interpreted as a new length 
scale that is derived from the model, and this length scale is not 
determined by any adjustable parameters in the problem other than the 
experimentally determined values 
of the Hubble constant and the background radiation temperature.  
The constraint on the position of the interface arises 
because the interface is a true shock--wave.  There is no similar constraint
on the position of the interface in the well-known Oppenheimer--Snyder model, 
where the interface is a contact discontinuity, \cite{oppesn,smolte1}.

In this paper we start with a critically expanding, ($k=0$), 
FRW metric under the assumption that the
equation of state agrees with the equation of state that applies in the
standard model of cosmology after the time of the thermal uncoupling of
matter from radiation.  This uncoupling occured at
a temperature of approximately $4000^oK,$ at about $300,000$ years after the
Big Bang in the standard model, \cite{blaugu,wein}.  In Section \ref{Sect3},
we derive a system of ODE's that determine the TOV metrics that match the
given FRW metric across a shock--wave interface, (equations (\ref{6.3}),
(\ref{6.4}) below). In fact, we derive the 
shock equations in the case of a general  FRW 
metric,
allowing for $k\neq0$ and for general equations of state, (equations (\ref{4.39}),
(\ref{4.40}) below).  In Section
\ref{Sect4} we give a derivation
of the FRW equation of state in terms of the cosmological scale factor.
This includes a discussion of the FRW metric in the presence of both
matter and radiation fields, assuming that the pressure 
due to matter is negligible, and that there is no 
thermal coupling between the fields. 
The results in Section \ref{Sect4} also apply for arbitrary $k.$  

To obtain the ODE's for the
TOV metric, we must rework the theory in \cite{smolte1,smolte3} where a given outer 
TOV metric is the starting point, instead of a given inner FRW metric which
we require here.  These ODE's, which are non-autonomous, simultaneously describe the TOV pressure 
$\bar{p}$ and the FRW shock position $r$, assuming conservation of energy and 
momentum and no delta function
sources at the shock. (We let barred quantities refer to TOV variables and
unbarred quantities to FRW variables, c.f. \cite{smolte1}.)  We then 
derive a formula for the TOV energy density
$\bar{\rho},$ (the only remaining undetermined variable in the TOV metric),  
which, together with solutions of the ODE's, determine the
TOV solutions that match the given FRW metric across a shock--wave 
interface.  
The ODE's take a particularly
simple form when the cosmological scale factor $R$ of the FRW metric 
is taken to be the
independent variable instead of the TOV radial variable $\bar{r},$ 
the independent variable in the usual formulation of the TOV system, 
\cite{wald,wein}.
In Section \ref{Sect5} we present a rather complete phase plane analysis of
these equations and we prove that there exists a unique bounded orbit.
This orbit describes the TOV pressure, but does not constrain either 
the initial shock position or the TOV energy density.  We show that
along this orbit, the pressure jump across the shock--wave has the 
property that the ratio of the TOV pressure to the FRW pressure at the
shock is equal to a slowly varying function that is bounded between
$1/9\approx.1111$ and $\bar{\sigma}=\sqrt{17}-4 \approx.1231,$ 
where the FRW pressure is supplied by the background radiation.  
Using this bound on the
TOV pressure along the orbit, we obtain the following sharp upper and 
lower bounds for
the (squared) distance that the shock--wave can propogate 
{\em over and above}
the (geodesic) motion of the galaxies, as a function of 
\lq\lq starting time'', $R_*.$
(Here $R=1$ denotes present time in the model, and we view the
starting time $R_*<1$ as the earliest time at which 
the shock--wave solution has settled down to the point where our model 
applies; that is, as entropy increases, we expect shock--wave solutions
to settle down to simple time asymptotic configurations, and we assume
here that this time asymptotic solution agrees with our model from $R_*$
onward.) The inequality reads, (see \ref{8.27a}) below),

$$
\frac{(2.62\times 10^{-7})T_0^4}{h_0^2H_0^2}
\ln\left(\frac{1}{R_*}\right)
\leq r^2-r_*^2\leq
\frac{(2.65\times 10^{-7})T_0^4}{h_0^2H_0^2}
\ln\left(\frac{1}{R_*}\right).
$$
Here the distance $r$ is given in terms of the Hubble length

$$
H_0^{-1}\approx \frac{.98}{h_0}\ \times10^{10}\ lightyears,
$$
where

$$
H_0=100h_0\ km\ sec^{-1}\ mpc^{-1},
$$
and it is generally agreed that $h_0$ lies in the interval
$.5\leq h_0\leq .85,$ \cite{peeb}.  
For example, if we take $T_0=2.736\approx 2.7^oK,$ $R_*=2.7/4000$, and $h_0=.55,$
(a recently quoted value), 
the above estimate reduces to

$$
r^2-r_*^2\approx\left(\frac{.019}{H_0}\right)^2.
$$ 
We conclude that
the distance the shock--wave has 
traveled, 
(over and above the motion of the galaxies),
between $R=R_*=2.7/4000$ and present time
$R=1,$  as predicted by this model,
is approximately $.019$ times the Hubble length.

In the standard interpretation of the FRW metric in Cosmology, 
the galaxies are in freefall, and
traverse geodesics $r=const.$  Thus we can interpret $r^2-r_*^2$ 
above as the (squared) distance that
the shock--wave travels over and above the motion due to freefall, 
a result of the fact that mass and momentum
are driven across the shock--wave as it evolves outward.  From this point
of view it is
a bit surprising that the quantity $r^2-r_*^2$ is independent of 
the starting position $r_*.$ 

Using the formula for the
TOV energy density, we next prove that the minimal physical requirement
$\rho>\bar{\rho}>\bar{p}>0,$ then places an additional contraint on the 
{\em initial} shock
position $r_*$ that depends on the starting time $R_*.$  
We prove that once this
constraint is met at one time, it is met at all succeeding times in the 
solution, and the density and pressure profiles are physically reasonable. 
Putting these results together, we obtain the following
upper and lower bounds on the shock position at the present time as a
function of background radiation temperature $T_0,$ the Hubble 
constant $H_0,$ and the value of the scale factor $R_*$ at which we
start the shock--wave, (c.f. (\ref{final1}) and
(\ref{final2}) below):

\begin{eqnarray}
r&\geq& H_0^{-1}\left\{(5.1\times10^{-4})\frac{T_0^2}{h_0}
\sqrt{\ln{\left(\frac{1}{R_*}\right)}}\right\},\nonumber\\
r&\leq& 
H_0^{-1}\sqrt{\frac{.76}
{1+\left[\frac{(4.6\times10^{-7})T_0^4}{h_0^2R_*^2}\right]}R_*+
(2.6\times10^{-7})\frac{T_0^4}{h_0^2}
\ln{\left(\frac{1}{R_*}\right)}}.\nonumber
\end{eqnarray}
The maximum shock position is plotted in Figure 2 for the case 
$T_0=2.7^oK,$
and $h_0=.55.$  In Section \ref{Sect8} we compare these bounds to the 
analagous bounds one obtains in the case of pure radiation, thus making
contact with the exact solution discussed in \cite{smolte2}.  For
example, at $T_0=2.7^oK$ we obtain

$$
\frac{36h_0}{H_0}\leq r\leq\frac{36h_0\sqrt{1+2.5R_*}}{H_0},
$$
c.f. (\ref{9.7}) below.  (Of course, since we are neglecting the matter field, 
we do not have 
$\dot{R}/R=H_0$ at the same time when $T=T_0$ in the pure radiation model.)
We note that
in this 
case the distance from the shock position
to the center of the explosion is {\em significantly} beyond the 
Hubble length.

In summary,  starting with the idea that there might be a shock--wave
that marks the outer boundary of the expansion that we measure by
the Hubble constant, one's first reaction is that nothing quantitative could be said
about the position of the shock without
knowing details concerning the nature of the spacetime beyond the 
shock--wave, or details about the mechanism that might have
created such an explosion in the first place.  And to a large extent
this must be true.  But what we
find interesting here is that this simplest shock--wave cosmological  
model, consistent with both the 
observed values of the Hubble constant and
the background radiation temperature, contains within it unexpected
constraints on the possible position of such a shock--wave, and the
shock position is comparable to the Hubble length.  That is,
we find it intriguing that this length scale comes out of the
model, and provides an answer that
apriori needn't have been so reasonable.  
   
In conclusion, we ask whether our expanding universe could have evolved 
from the center of a great 
explosion that generated a shock--wave at its leading edge.  If so it makes
sense to wonder whether some of 
the 
far away objects that we observe in the nightime sky are possibly due to 
similar explosions that originated at
other locations in spacetime.  We now know that the scale of supernovae is 
not the largest scale on which classical explosions 
have occured in the universe.  Indeed, it was reported in a 
recent issue of Nature, that  
on May 7, 1998, a gamma ray  explosion emanating from a faint galaxy known 
as GRB971214 erupted, and for two seconds the burst was more luminous 
than the rest of the 
universe
combined. This is the largest explosion ever recorded, and redshifts 
place it at about 12 billion lightyears away. Moreover, conditions at 
the explosion were equivalent to those one millisecond after the 
Big Bang in 
the standard model.
Thus we pose the question: could explosions such as this, 
or even greater than this,
 have given rise to our own ``expanding universe''?  
Indeed, could we then
observe other similar explosions in distant
regions of 
spacetime beyond the expansion of our own universe, (that is, beyond the 
shock--wave that marks the edge of the expansion we measure by the Hubble
constant)?  We propose the shock--wave model presented in this paper as a natural and simple starting point
for a futher investigation of these issues.  But independently of this, the model provides a new, 
essentially {\em exact} solution of the Einstein equations
that we feel
is interesting in its own right.

\section{Preliminaries}\label{Sect2}
\setcounter{equation}{0}

According to Einstein's theory of general relativity, \cite{eins}, the
gravitational field is described by a Lorenzian metric $g$ that satisfies the Einstein equations

\begin{equation}
\label{2.1}
G=\frac{8\pi{\cal G}}{c^4}T,
\end{equation}
on $4$-dimensional spacetime.
Here $G$ is the Einstein curvature tensor, ${\cal G}$ denotes
Newton's gravitational constant, $c$ denotes the speed of light,
and $T$ is the stress energy
tensor, the source of the gravitational field.  In this paper 
we are concerned with FRW and TOV metrics, two spherically symmetric
metrics which are
exact solutions of (\ref{2.1})
when $T$ takes the form of a stress tensor for a perfect
fluid, namely

\begin{equation}
\label{2.2}
T_{ij}=(p+\rho c^2)u_iu_j+pg_{ij},
\end{equation} 
where $\rho$ denotes the mass-energy density, $p$ the pressure and 
$i,j=0,...,3$ denote indices of spacetime coordinates.  
The FRW metric is given by

\begin{equation}
\label{2.3}
ds^2=-d(ct)^2+R^2(t)\left\{\frac{1}{1-kr^2}dr^2+r^2d\Omega^2\right\},
\end{equation}
and the TOV metric is given by

\begin{equation}
\label{2.3a}
d\bar{s}^2=-B(\bar{r})d(c\bar{t})^2+A(\bar{r})^{-1}d\bar{r}^2+
\bar{r}^2d\Omega^2,
\end{equation}
where $d\Omega^2$ denotes the standard metric on the unit two-sphere.
We write the TOV metric in barred coordinates so that it can be 
distinguished from the unbarred FRW coordinates when we do the matching
of these two metrics below, c.f. \cite{smolte1}.  Assuming co-moving
coordinates and substituting (\ref{2.3}) into the (\ref{2.1})
yields the following FRW equations, \cite{misnthwh,peeb,wald,wein}: 

\begin{equation}
\label{2.3b}
\dot{R}^2=\frac{8\pi{\cal G}}{3c^4}\rho R^2-k,
\end{equation}
and

\begin{equation}
\label{2.3c}
p=-\rho-\frac{R\dot{\rho}}{3\dot{R}}.
\end{equation}
The unknowns $R,$ $\rho$ and $p$ in the FRW equations are 
assumed to functions of the FRW time $t$ alone, and \lq\lq dot''  
denotes 
differentiation with respect to $t.$   Assuming co-moving
coordinates and substituting (\ref{2.3a}) into the (\ref{2.1})
yields the following TOV equations,

\begin{equation}
\label{2.4}
\frac{dM}{d\bar{r}}=4\pi\bar{r}^2\bar{\rho},
\end{equation}

\begin{equation}
\label{2.5}
-\bar{r}^2\frac{d\bar{p}}{d\bar{r}}={\cal G}M\bar{\rho}
\left(1+\frac{\bar{p}}{\bar{\rho}}\right)
\left(1+\frac{4\pi\bar{r}^3\bar{p}}{M}\right)A^{-1},
\end{equation}
and

\begin{equation}
\label{2.6}
\frac{B'}{B}=-2\frac{\bar{p}'}{\bar{p}+\bar{\rho}},
\end{equation}
where
\begin{equation}
\label{2.7}
A=1-\frac{2{\cal G}M}{c^2\bar{r}}.
\end{equation}
Here the unknown functions are the density $\bar{\rho},$ the pressure $\bar{p},$
and the total mass $M,$ \cite{schoya},
which are assumed to be functions of $\bar{r}$ alone, and prime
denotes differentiation with respect to $\bar{r}.$  In the next section
we fix an FRW metric and derive equations for the TOV metrics that
match the given FRW metric across a shock--wave interface at which
the metric is only Lipschitz continuous, and across which conservation
of mass and momentum hold, and at which there are no delta function
sources. In \cite{smolte1} it is shown that the shock surface is 
given implicitly by

\begin{equation}
\label{2.8}
M(\bar{r})=\frac{4\pi}{3}\rho(t)\bar{r}^3,
\end{equation}
and the metrics (\ref{2.3}), (\ref{2.3a}) are identified
via a coordinate transformation in which

\begin{equation}
\label{2.9}
\bar{r}=Rr.
\end{equation}

\section{Derivation of Equations}
\label{Sect3}
\setcounter{equation}{0}

In this section we derive equations that describe the time evolution of
an outgoing
spherical shock--wave interface together with an outer TOV metric, 
such that the
shock surface matches a given FRW metric on the inside, and such that conservation of
energy and momentum hold across the interface.  The main point here is 
that we are assuming a given {\em inner} FRW metric, rather than assuming 
a given {\em outer} TOV metric as in \cite{smolte1,smolte3}.  Thus we seek a pair of 
equations that
determine an outer TOV metric that matches a given FRW across a shock--wave
interface.  Rather than deriving the shock equations, we shall write them
down and prove that solutions of these equations determine a shock--wave
solution of the Einstein equations.  (The reader can obtain a formal
derivation of these equations by reversing the steps in the arguments 
below.)

Equation (\ref{4.8}) in our first theorem below is the first equation in
the pair of ODE's that we will work with.

\begin{Theorem}\label{T4.1}
Assume that $\rho(t),$ $p(t),$ and $R(t)$ solve the FRW system

\begin{equation}
\label{4.1}
\dot{R}=\sqrt{\frac{8\pi{\cal G}}{3c^4}R^2\rho-k},
\end{equation}

\begin{equation}
\label{4.2}
\dot{\rho}=-3\frac{\dot{R}}{R}(\rho+p),
\end{equation}
over some interval

\begin{equation}
\label{4.3}
I=(t_1,t_2).
\end{equation}
Assume that 

\begin{equation}
\label{4.3b1}
R(t)>0,
\end{equation}
and that 

\begin{equation}
\label{4.3b}
\dot{R}\neq0,
\end{equation}
on $I.$  We assume WLOG, (by the
choice of positive square root in (\ref{4.1})), that

\begin{equation}
\label{4.4}
\dot{R}>0.
\end{equation}
Assume further that $r(t)$ is a positive invertible function defined on 
$I,$ and define $\bar{r}(t)$ on $I$ by

\begin{equation}
\label{4.5}
\bar{r}=Rr.
\end{equation}
Define functions $M(\bar{r})$ and $\bar{\rho}(\bar{r}))$ by

\begin{equation}
\label{4.6}
M(\bar{r}(t))=\frac{4\pi}{3}\rho(t)\bar{r}(t)^3,
\end{equation}
and 

\begin{equation}
\label{4.7}
\bar{\rho}(\bar{r})=\frac{M'(\bar{r})}{4\pi\bar{r}^2},
\end{equation}
where prime denotes differentiation with respect to $\bar{r}.$   
Assume, finally, that $r(t)$ satisfies
  
\begin{equation}
\label{4.8}
\dot{r}=\frac{1}{R}\left(\frac{p-\bar{p}}{\rho+\bar{p}}\right)
\frac{1-kr^2}{\dot{R}r},
\end{equation}
for some function $\bar{p},$  and that $\rho,$ $p,$ $M,$ $\bar{\rho},$ 
and $\bar{p}$ are all positive valued functions
on $I.$  Then for all $t\in I$ we have,

\begin{equation}
\label{4.9}
p=\frac{\gamma\theta\bar{\rho}-\rho}{1-\gamma\theta},
\end{equation}
where, \cite{smolte3},

\begin{equation}
\label{4.10}
\theta=\frac{A}{1-kr^2},
\end{equation}

\begin{equation}
\label{4.11}
A=1-\frac{2{\cal G}M}{c^4\bar{r}},
\end{equation}
and

\begin{equation}
\label{4.12}
\gamma=\frac{\rho+\bar{p}}{\bar{\rho}+\bar{p}}.
\end{equation}
\end{Theorem}
That is, Theorem 1 implies that for a given FRW solution, (\ref{4.8})
implies the conservation condition (\ref{4.9}) when $M$ and $\bar{\rho}$
are defined by (\ref{4.6}), and (\ref{4.7}), (these latter two equations being the shock surface
matching condition and the second TOV equation, respectively, 
\cite{smolte1,smolte3}).  Here dot denotes
$\frac{d}{ct},$ and we assume $c=1.$

\vspace{.3cm}
\noindent {\bf Proof:}  Differentiating  (\ref{4.6}) with respect to
$ct$ and using (\ref{4.7}) gives

\begin{equation}
\label{4.13}
\dot{M}=\frac{dM}{d\bar{r}}\dot{\bar{r}}=
4\pi\bar{\rho}\bar{r}^2\dot{\bar{r}}.
\end{equation}
But (\ref{4.6}) gives

\begin{equation}
\label{4.14}
\dot{M}=\frac{4\pi}{3}\dot{\rho}\bar{r}^3+4\pi\rho\bar{r}^2\dot{\bar{r}},
\end{equation}
so from (\ref{4.13}) and (\ref{4.14}) we get

\begin{equation}
\label{4.15}
\dot{\bar{r}}=\frac{Rr}{3(\bar{\rho}-\rho)}\dot{\rho}.
\end{equation}
Using (\ref{4.2}) in (\ref{4.15}) gives

\begin{equation}
\label{4.16}
\dot{\bar{r}}=-r\dot{R}\frac{p+\rho}{\bar{\rho}-\rho}.
\end{equation}
Using (\ref{4.5}) and simplifying we have

\begin{equation}
\label{4.17}
\dot{r}R+r\dot{R}=-\dot{R}r\left(\frac{\rho+p}{\bar{\rho}-\rho}\right).
\end{equation}
Using (\ref{4.8}) to eliminate $\dot{r}$ from (\ref{4.17}) gives

\begin{equation}
\label{4.18}
\frac{1-kr^2}{\dot{R}^2r^2}=-\left(\frac{p+\bar{\rho}}{\bar{\rho}-\rho}
\right)
\left(\frac{\rho+\bar{p}}{p-\bar{p}}\right).
\end{equation}
We now use the identity 

\begin{equation}
\label{4.19}
\frac{1-kr^2}{\dot{R}^2r^2}=\frac{1}{1-\theta},
\end{equation}
which follows from (\ref{4.1}) and (\ref{4.6}).  Indeed, 

$$
\dot{R}^2=\frac{2{\cal G}}{c^4}\frac{M}{\bar{r}^3}R^2-k.
$$
But (\ref{4.11}) implies that

$$
\frac{2{\cal G}M}{c^4}=(1-A)\bar{r},
$$
and using this gives

$$
\dot{R}^2=\frac{1-A}{r^2}-k,
$$
or, (c.f. \cite{smolte1,smolte3}),

\begin{equation}
\label{4.20}
r^2\dot{R}^2=-A+(1-kr^2).
\end{equation}
Using (\ref{4.10}) in (\ref{4.20}) gives (\ref{4.19}), as claimed.  

Now using
(\ref{4.19}) in (\ref{4.18}) yields

$$
\frac{1}{1-\theta}=-\left(\frac{p+\bar{\rho}}{\bar{\rho}-\rho}\right)
\left(\frac{\rho+\bar{p}}{p-\bar{p}}\right).
$$
Solving this for $p$ gives (\ref{4.9}), where we have used (\ref{4.12}).
This completes the proof of Theorem 1.

For a given FRW metric, Theorem 1 tells us that the ODE (\ref{4.8}) can
be taken in place of the conservation constraint (\ref{4.9}), and the 
reversal of the steps in the above proof can be regarded as a formal
derivation of the ODE (\ref{4.8}).  For later convenience, we now record
the following additional equations that follow from the hypotheses of 
Theorem \ref{4.1}.

\begin{Corollary}\label{C4.2}

Assume that the hypotheses (\ref{4.1}) through
(\ref{4.8}) of Theorem \ref{4.1} hold.  Then the following equations
are valid:

\begin{equation}
\label{4.21}
\frac{\rho+p}{\rho-\bar{\rho}}=\frac{\gamma\theta}{\gamma\theta-1},
\end{equation}

\begin{equation}
\label{4.22}
\dot{\bar{r}}(\bar{\rho}+\bar{p})=\sqrt{1-kr^2}
\left(\frac{\theta}{\sqrt{1-\theta}}\right)(p-\bar{p}),
\end{equation}

\begin{equation}
\label{4.23}
\frac{\theta}{1-\theta}=\left(\frac{\rho+p}{\bar{\rho}-\rho}\right)
\left(\frac{\bar{\rho}+\bar{p}}{p-\bar{p}}\right),
\end{equation}

\begin{equation}
\label{4.24}
\dot{\bar{r}}=\frac{\gamma\theta}{\gamma\theta-1}\sqrt{1-kr^2}
\sqrt{1-\theta},
\end{equation}

\begin{equation}
\label{4.25}
r^2\dot{R}^2=-A+\left(1-kr^2\right),
\end{equation}

\begin{equation}
\label{4.26}
\frac{1-kr^2}{r^2\dot{R}^2}=\frac{1}{1-\theta}.
\end{equation}
\end{Corollary}  

\vspace{.3cm}
\noindent {\bf Proof:} By Theorem 4.1, we know that (\ref{4.9}) holds,
and using this in the LHS of (\ref{4.21}) gives the RHS of (\ref{4.21}).
Also, from (\ref{4.20}), 

\begin{equation}
\label{4.27}
r\dot{R}=\sqrt{1-kr^2}\sqrt{1-\theta}, 
\end{equation}
and using this in (\ref{4.17}) gives

\begin{equation}
\label{4.28}
\dot{\bar{r}}=-\sqrt{1-kr^2}\sqrt{1-\theta}
\left(\frac{\rho+p}{\bar{\rho}-\rho}\right).
\end{equation}
Using (\ref{4.21}) in (\ref{4.28}) gives (\ref{4.24}).  From (\ref{4.9})
we get

$$
p-\bar{p}=\frac{\gamma\theta\bar{\rho}-\rho}{1-\gamma\theta}-
\frac{\bar{p}(1-\gamma\theta)}{1-
\gamma\theta}=\frac{(\rho+\bar{p})\theta)-(\rho+\bar{p})}{1-
\gamma\theta},
$$
so

\begin{equation}
\label{4.29}
\frac{p-\bar{p}}{\rho+\bar{p}}=\frac{\theta-1}{1-\gamma\theta}.
\end{equation}
To verify (\ref{4.22}), we use (\ref{4.24}) which we write in the form
$$
\dot{\bar{r}}\sqrt{1-kr^2}\frac{1}{\sqrt{1-\theta}}\frac{1-
\theta}{\gamma\theta-1}\theta,
$$
and so from (\ref{4.29}) we have

\begin{equation}
\label{4.30}
\dot{\bar{r}}=\frac{\sqrt{1-kr^2}}{\sqrt{1-\theta}}\theta
\left(\frac{p-\bar{p}}{\bar{\rho}-\bar{p}}\right).
\end{equation}
Solving for ($\bar{\rho}+\bar{p}$)$\dot{\bar{r}}$ in (\ref{4.30}) gives
(\ref{4.22}).  Finally, to obtain (\ref{4.23}), equate the RHS's of
(\ref{4.28}) and (\ref{4.30}).  Equations (\ref{4.25}) and (\ref{4.26})
have already been derived as (\ref{4.20}) and (\ref{4.19}) within the 
proof of Theorem \ref{T4.1}.  This completes the proof of the Corollary.

Now assume that $\rho(t),$ $p(t),$ and $R(t)$ solve the FRW system
(\ref{4.1}) and (\ref{4.2}) for $t\in I,$ and assume that the hypotheses
(\ref{4.3}) to (\ref{4.8}) of Theorem \ref{T4.1} hold.  We know from Theorem
\ref{T4.1} that the conservation condition (\ref{4.9}) also holds.  We
now find an equation for $\bar{p}(\bar{r}),$ (equation (\ref{4.34}) below), which guarantees that
$\bar{p}$ solves the TOV equation (\ref{2.5}), since then, in light
of (\ref{4.7}), the functions $\bar{\rho}(\bar{r}),$ $\bar{p}(\bar{r}),$
and $M(\bar{r})$ will then solve the TOV system as well.  Defining
$A(\bar{r})$ by (\ref{4.11}), we can define the function $B(\bar{r})$ as
a solution of the ODE, \cite{smolte1,smolte3},

\begin{equation}
\label{4.31}
\frac{B'(\bar{r})}{B(\bar{r})}=-
\frac{2\bar{p}'(\bar{r})}{\bar{\rho}(\bar{r})+\bar{p}(\bar{r})},
\end{equation}
thus determining a TOV metric of the form

\begin{equation}
\label{4.32}
ds^2=-B(\bar{r})d\bar{t}^2+A^{-1}(\bar{r})d\bar{r}^2+\bar{r}^2d\Omega^2,
\end{equation}
that solves the Einstein equation $G=\kappa T$ for a perfect fluid with stress tensor

$$
T_{ij}=\bar{p}g_{ij}+(\bar{\rho}+\bar{p})u_iu_j.
$$
For this metric, co-moving coordinates are assumed, \cite{wein}, and thus the $4$-velocity ${\bf u}$ is given by

$$
u^0=\sqrt{B},\ \ \ u^i=0,\  i=1,2,3.
$$
Note that we are free to choose any positive initial value for $B$ by suitable
rescaling of the time coordinate $\bar{t}.$  The next 
lemma demonstrates that if
$\bar{p}$ satisfies equation (\ref{4.34}) below, then as a consequence it
also satisfies

\begin{equation}
\dot{\bar{p}}=-\frac{{\cal G}M\dot{\bar{r}}}{c^4\bar{r}^2}
\bar{\rho+\bar{p}}\left(1+\frac{4\pi\bar{p}\bar{r}^3}{M}\right)A^{-1},
\label{4.33}
\end{equation}
which is equivalent to the TOV equation (\ref{2.5}).

\begin{Lemma}\label{L4.3}
The hypotheses (\ref{4.1}) to (\ref{4.8}) of Theorem
\ref{4.1}, together with the equation

\begin{equation}
\label{4.34}
\dot{\bar{p}}=-\frac{{\cal G}M}{c^4\bar{r}^2}
\left(\frac{1-kr^2}{\dot{R}r}\right)
\left(\frac{A}{1-kr^2}\right)
(p-\bar{p})\left(1+3\frac{\bar{p}}{\rho}\right)A^{-1},
\end{equation}
imply that $\bar{p}(\bar{r})$ also solves the TOV equation (\ref{2.5}).
\end{Lemma}

\vspace{.3cm}
\noindent{\bf Proof:}  By (\ref{4.26}) we have

$$
\frac{\sqrt{1-kr^2}}{\dot{R}r}=\frac{1}{\sqrt{1-\theta}},
$$
and substituting this together with (\ref{4.10}) into (\ref{4.34}) gives 

\begin{equation}
\label{4.35}
\dot{p}=-\frac{{\cal G}M}{c^4\bar{r}^2}
\sqrt{1-kr^2}\left(\frac{\theta}{\sqrt{1-\theta}}\right)(p-\bar{p})
\left(1+3\frac{\bar{p}}{\rho}\right)A^{-1}.
\end{equation}
But using (\ref{4.22}) and (\ref{4.6}) in (\ref{4.35}) we obtain

\begin{equation}
\label{4.36}
\dot{\bar{p}}=
-\frac{{\cal G}M}{c^4\bar{r}^2}\dot{\bar{r}}(\bar{\rho}+\bar{p})
\left(1+\frac{4\pi\bar{p}\bar{r}^3}{\frac{4\pi}{3}\rho\bar{r}^3}\right)A^{-1},
\end{equation}
which directly implies the TOV equation (\ref{2.5}).  This completes
the proof of the Lemma \ref{L4.3}.

Our results now imply the following theorem which introduces the system
of ODE's whose solutions we analyze in subsequent sections:

\begin{Theorem}\label{T4.4}  Assume that $\rho(t),$ $p(t),$ and $R(t)$
satisfy the FRW equations (\ref{4.1}) and (\ref{4.2}) for $t\in I,$
and that the other
hypotheses (\ref{4.3}) through (\ref{4.8}) of Theorem \ref{4.1} hold.
Assume further that ($r(t),\bar{p}(t)$) solves the system of ODE's

\begin{eqnarray}
\dot{r}&=&\frac{1}{R}\left(\frac{p-\bar{p}}{\rho+\bar{p}}\right)
\frac{1-kr^2}{\dot{R}r},\label{4.37}\\ \nonumber \\
\dot{\bar{p}}&=&-\frac{{\cal G}M}{c^4\bar{r}^2}
\frac{(p-\bar{p})\left(1+3\frac{\bar{p}}{\rho}\right)}{\dot{R}r},\label{4.38}
\end{eqnarray}
for $t\in I,$ where $\bar{r},$ $M(\bar{r}),$ and $\bar{\rho}(\bar{r})$
are defined for $\bar{r}(t_1)<\bar{r}<\bar{r}(t_2)$ by (\ref{4.5}),
(\ref{4.6}) and (\ref{4.7}).  Then $\bar{\rho}(\bar{r}),$ 
$\bar{p}(\bar{r}),$
$M(\bar{r}),$ solve the TOV system (\ref{2.4}), (\ref{2.5}), and the conservation 
condition (\ref{4.9}) holds for all $t\in I.$  Furthermore, under these assumptions, 
the system (\ref{4.37}),
(\ref{4.38}) is equivalent to the system

\begin{eqnarray}
\frac{dr}{dR}&=&\frac{1}{R(QR^2-k)}\left(\frac{P-\bar{P}}
{Q+\bar{P}}\right)\left(\frac{1-kr^2}{r}\right),\label{4.39}\\\nonumber\\
\frac{d\bar{P}}{dR}&=&-\frac{1}{2}R\frac{(Q+3\bar{P})(P-\bar{P})}
{QR^2-k},\label{4.40}
\end{eqnarray} 
for $R(t_1)<R<R(t_2),$ where 

\begin{equation}
\label{4.41}
(Q,P,\bar{P})=\frac{8\pi{\cal G}}{3c^4}(\rho,p,\bar{p}),
\end{equation}
has the dimensions of inverse length squared.  
\end{Theorem}

Note that the equivalence of system (\ref{4.37}), (\ref{4.38}) with 
(\ref{4.39}), (\ref{4.40}) follows because of the assumption 
$\dot{R}\neq0.$  This also  
implies that $P$ and $Q$ can
be considered as functions of $R,$ in which case equations (\ref{4.39})
and (\ref{4.40}) closes to form a well-defined nonlinear system of two
ODE's in the unknowns $r$ and $\bar{P}.$  After solving (\ref{4.39}), (\ref{4.40})
the dependence of $R$ on $t$ can be recovered from (\ref{4.1}).  
Thus for a given FRW metric and a given solution of (\ref{4.39}), (\ref{4.40}), the only
variable remaining to be determined is the TOV energy density $\bar{Q}$ on the outside of 
the (outgoing) shock--wave.  To obtain a closed form expression for $\bar{Q},$ write
the shock surface equation (\ref{4.6}) in the form

\begin{equation}
\label{fill1}
2{\cal G}M=Q\bar{r}^3.
\end{equation}
But the second TOV equation for $M$ is

\begin{equation}
\label{fill2}
\frac{dM}{d\bar{r}}=4\pi\bar{\rho}\bar{r}^2,
\end{equation}
which we can rewrite in the form

\begin{equation}
\label{fill3}
\frac{d}{d\bar{r}}(2{\cal G}M)=3\bar{Q}\bar{r}^2.
\end{equation}
Differentiating (\ref{fill1}), substituting into (\ref{fill3}) and solving for $\bar{Q}$
yields the formula

\begin{equation}
\label{fill4}
\bar{Q}=\frac{1}{3\bar{r}^2}\frac{d}{d\bar{r}}(Q\bar{r}^3)=Q+\frac{Rr}{3}\frac{d}{d\bar{r}}Q.
\end{equation}
Note that if $Q$ decreases as the shock moves outward, (that is, the $\bar{r}$ position of the
shock increases), then the second term in (\ref{fill4}) is negative, and so $\bar{Q}<Q,$ (the 
density behind the shock is greater than the density in front of the shock), as is the case for
classical shock--waves in fluids, \cite{smol}.  Note, however that the physically
necessary condition $\bar{Q}>0,$ or the physically reasonable condition $\bar{Q}>\bar{P},$ is not guaranteed, 
and depends on the particular solution.

The final theorem of this section tells us that solutions of the ODE's
(\ref{4.38}), (\ref{4.39}) do indeed determine exact shock--wave solutions of the 
Einstein equations when a (suitable) FRW metric is given.

\begin{Theorem}\label{T4.5}  Assume that $\rho(t),$ $p(t),$ and $R(t)$
satisfy the FRW equations (\ref{4.1}) and (\ref{4.2}) for $t\in I,$
and that the 
hypotheses (\ref{4.3}) through (\ref{4.7}) of Theorem \ref{4.1} hold.
Assume further that ($r(R),\bar{P}(R)$) solve the system of ODE's 
(\ref{4.39}), (\ref{4.40}) for $R(t_1)<R<R(t_2).$  Assume that $Q,$ $P,$ 
$M,$ $\bar{Q},$ 
$\bar{P},$ and $A$ are all positive and that the shock
speed in FRW coordinates is less than the speed of light throughout 
the interval $I.$  Then there exists a $C^{1,1}$ invertible coordinate 
transformation mapping $(t,r)\rightarrow(\bar{t},\bar{r})$ of the form

\begin{eqnarray}
\label{4.42}
\bar{t}&=&\bar{t}(t,r),\\
\bar{r}&=&\bar{r}(t,r)\equiv R(t)r,
\end{eqnarray}
such that, under this identification, the resulting TOV metric matches the
given FRW metric Lipschitz continuously across the shock surface $r=r(t).$  (The
angular coordinates $\theta$ and $\phi$ are implicitly identified.)
Moreover, the Lipschitz continuous metric defined by taking the FRW metric
for $r<r(t)$ and the TOV metric for $r>r(t)$ defines a shock--wave solution
of the Einstein equations (\cite{isra,misnthwh}), c.f. 
\cite{smolte1,smolte3}.
In particular, the Rankine-Hugoniot jump conditions 

\begin{equation}
\label{4.43}
[T_{ij}]n^i=0,\ \ j=0,...,3,
\end{equation}
hold across the shock surface; there are no \lq\lq delta function sources"
on the surface; there exists a regular $C^{1,1}$ coordinate transformation
defined in a neighborhood of each point on the shock such that the metric
components in the transformed coordinates, (which can be taken to be Gaussian
normal coordinates), have smoothness level $C^{1,1}$; and the
matched metric determines a weak solution of the Einstein equations in the sense of the theory of
distributions, c.f. \cite{smolte1}.
\end{Theorem}

\vspace{.3cm}
\noindent{\bf Proof:} The existence of the coordinate transformation is
proved in \cite{smolte1} pages 278-280 under the assumption that
the shock surface is nowhere characteristic in the sense of (4.43) of that
paper.  Using (4.55) of the same reference, the non-characteristic 
condition
can be re-written as

\begin{equation}
\label{4.44}
\dot{\bar{r}}\neq-\frac{A}{\dot{R}r},
\end{equation}
which holds here because we assume that $A>0,$ $r>0,$ $\dot{R}>0$ and
$\dot{\bar{r}}>0.$
Since the normal vector ${\bf n}$ to the shock surface is non-null, 
(because
we assume that the shock speed is less than the speed of light), and
the functions $c(t,r)$ and $\bar{c}(\bar{t},\bar{r})$ in Lemma 9 of
\cite{smolte1} are here equal to $R(t)r$ and $\bar{r},$ 
respectively, it follows that the conclusions of Lemma 9, 
\cite{smolte1}
are valid.  Moreover, the conservation condition (\ref{4.9}) is valid, and
thus the argument in \cite{smolte1} that leads to (3.9) in this 
latter reference, implies that, under our hypotheses, 
condition (5.5) of Lemma 9, 
\cite{smolte1}, 
follows from the conservation condition (\ref{4.9}) above.  (Note that the 
condition (2.20) of \cite{smolte1}, assumed in that paper, is not 
needed
here.)  Since the
conclusions of Theorem \ref{T4.5} are just a re-statement of the 
conclusions
of Lemma 9, \cite{smolte1}, the proof of Theorem \ref{T4.5} is now
complete.
\vspace{.2cm}

A remarkable aspect of the formulation of the shock equations given 
in (\ref{4.39}) and
(\ref{4.40}) is that, if $Q$ and $P$ are given functions of $R,$ 
(which can be obtained from the FRW equations
once an equation of state is specified), then the equation 
(\ref{4.40}) for $\bar{P}$ {\em uncouples} from the 
$\bar{r}$ equation (\ref{4.39}).
Thus, in principle, one can solve system (\ref{4.39}) and (\ref{4.40}) 
by first solving the scalar non-autonomous 
equation
(\ref{4.40}) for $\bar{P},$ and then plugging the solution $\bar{P}(R)$ 
into (\ref{4.39}) to obtain a 
scalar non-autonomous ODE for
the shock position $r.$

As an application, and to clarify the way system (\ref{4.39}) and (\ref{4.40}) works, we now
demonstrate that the set of shock--wave solutions of the Einstein equations determined by system
(\ref{4.39}), (\ref{4.40}) includes, as a special case, the exact solutions first presented in 
\cite{smolte2}.   
To verify this, set  $k=0$ in (\ref{4.39}) and (\ref{4.40}) to obtain the system

\begin{eqnarray}
\frac{dr}{dR}&=&\frac{1}{QR^3}\left(\frac{P-\bar{P}}
{Q+\bar{P}}\right)\left(\frac{1}{r}\right),\label{4.39.a}\\\nonumber\\
\frac{d\bar{P}}{dR}&=&-\frac{1}{2}\frac{(Q+3\bar{P})(P-\bar{P})}
{QR}.\label{4.40.b}
\end{eqnarray}
Now it is easy to verify that the solution in \cite{smolte2}, (as given in equations (5.1) through (5.14) of
that paper), satisfies the following equations:
\begin{equation}
P=\sigma Q,
\label{4.a}
\end{equation}

\begin{equation}
\bar{P}=\frac{\bar{\sigma}}{3}\bar{Q},
\label{4.b}
\end{equation}
for some constants $0<\sigma,\bar{\sigma}<1$, and

\begin{equation}
\frac{Q}{Q_0}=\left(\frac{R}{R_0}\right)^{-3(1+\sigma)},
\label{4.c}
\end{equation}

\begin{equation}
\label{4.d}
\left(\frac{r}{r_0}\right)^2=\left(\frac{R}{R_0}\right)^{1+3\sigma},
\end{equation}
where the subscript zero denotes values at some particular time, say
present time in the FRW solution.  For our purposes here, we now take 
(\ref{4.a}) to
(\ref{4.d}) as an ansatz, and show that this ansatz is consistent with system (\ref{4.39.a}), (\ref{4.40.b}), 
and that system (\ref{4.39.a}) and (\ref{4.40.b}) then determines the other relations in (5.1) to
(5.14) of \cite{smolte2}, (including the relation between $\sigma$ and $\bar{\sigma}$).   To this end, set 
$R_0=1$\footnotemark[3]\footnotetext[3]{Note that we are free to fix the initial condition $R_0=1$ because the 
FRW metric is invariant under 
rescalings of $R$ when $k=0.$  Indeed, under the scaling $R\rightarrow aR,$ the 
variables $Q,$ $P,$ $\bar{P}$ and $\bar{r}$ are invariant, so the shock position $r=\bar{r}/R$ rescales like 
$r\rightarrow a^{-1}r.$  It is easy to see that system (\ref{4.39.a}), (\ref{4.40.b}) is invariant under this 
rescaling.}  
 and $c=1$, and substitute (\ref{4.a})-(\ref{4.c}) into equation 
(\ref{4.40.b}) to obtain 

\begin{equation}
\frac{R}{Q}\frac{dQ}{dR}=-\frac{3}{2}\frac{(1+\bar{\sigma})(\sigma-
\bar{\sigma})}{\bar{\sigma}}.
\label{4.e}
\end{equation} 
But from (\ref{4.c}),

\begin{equation}
\frac{R}{Q}\frac{dQ}{dR}=\frac{d\ln(Q)}{d\ln(R)}=-3(1+\sigma),
\label{4.f}
\end{equation}
and putting this into the left hand side of (\ref{4.e}) and solving for $\sigma$ gives

\begin{equation}
\sigma=\frac{\bar{\sigma}(7+\bar{\sigma})}{3(1-\bar{\sigma})},
\label{4.g}
\end{equation}
which is the relation between $\sigma$ and $\bar{\sigma}$ given in (\ref{5.2})-(\ref{5.3}) of
\cite{smolte2}.

Substituting (\ref{4.a}) and (\ref{4.b}) into equation (\ref{4.39.a}) and simplifying gives

\begin{equation}
\frac{1}{2}\frac{dr^2}{dR}=\frac{1}{QR^2}\left(\frac{3\sigma-\bar{\sigma}}{3+\bar{\sigma}}\right),
\label{4.h}
\end{equation} 
and using (\ref{4.c}) and (\ref{4.d}) in (\ref{4.h}) gives, after simplification,

\begin{equation}
Q_0r_0^2=2\frac{3\sigma-\bar{\sigma}}{(1+3\sigma)(3+\bar{\sigma})}.
\label{4.i}
\end{equation}
Equation (\ref{4.i}) determines the constant $r_0$ from the constants $Q_0$ and $\sigma$ or
$\bar{\sigma}.$  To connect this with the formulas in \cite{smolte2}, note that (\ref{4.c}) and
(\ref{4.d}) together with the matching condition

\begin{equation}
\bar{r}=Rr,
\label{4.j}
\end{equation}
yeild the formula

\begin{equation}
Q=Q_0\left(\frac{\bar{r}}{\bar{r}_0}\right)^2.
\label{4.k}
\end{equation}
Since we take $R_0=1,$ 

\begin{equation}
Q=\frac{Q_0r_0^2}{\bar{r}_0^2}.
\label{4.l}
\end{equation}
But (\ref{4.l}) checks with \cite{smolte2} because (3.2)-(3.4) there imply that

\begin{equation}
Q=\frac{8\pi{\cal G}}{3}\rho=\frac{8\pi{\cal G}\gamma}{\bar{r}^2},
\label{4.m}
\end{equation}
where, (using the notation of \cite{smolte2}),

\begin{equation}
\gamma=\frac{1}{2\pi{\cal G}}\left(\frac{\bar{\sigma}}{1+6\bar{\sigma}+\bar{\sigma}^2}\right).
\label{4.n}
\end{equation}
A calculation using (\ref{4.g}) verifies that

\begin{equation}
2\frac{3\sigma-\bar{\sigma}}{(1+3\sigma)(3+\bar{\sigma})}=8\pi{\cal G}\gamma,
\label{4.o}
\end{equation}
and thus (\ref{4.i}) implies (\ref{4.m}) as claimed.

We can now solve for the density $\bar{Q}$ using the TOV equation
\begin{equation}
\frac{dM}{d\bar{r}}=4\pi\bar{\rho}\bar{r}^2,
\label{4.p}
\end{equation}
together with the shock matching condition

\begin{equation}
M=\frac{4\pi{\cal G}}{3}\rho\bar{r}^3.
\label{4.q}
\end{equation}
That is, using (\ref{4.l}) in (\ref{4.q}) gives

\begin{equation}
M(\bar{r})=4\pi\gamma\bar{r},
\label{4.r}
\end{equation}
and thus (\ref{4.p}) implies

\begin{equation}
\bar{\rho}=\frac{\gamma}{\bar{r}^2},
\label{4.s}
\end{equation}
which implies that

\begin{equation}
\bar{Q}=\frac{1}{3}Q,
\label{4.t}
\end{equation}
and

\begin{equation}
\bar{P}=\bar{\sigma}\bar{Q},
\label{4.u}
\end{equation}
also agreeing with the formulas arrived at in \cite{smolte2}.  Finally, substituting
(\ref{4.c}) into the FRW equation

\begin{equation}
\dot{R}^2=QR^2,
\label{4.v}
\end{equation}
and integrating gives the time dependence of $R$ as

\begin{equation}
R(t)=R_0\left[\frac{(2+3\sigma)\sqrt{Q_0}}{2}t\right]^{\frac{2}{2+3\sigma}},
\label{4.w}
\end{equation}
where we have used the initial condition $R(0)=0.$

For our construction below, the case $\sigma=\frac{1}{3}$ is relevant, (the case of pure
radiation).
In this case,

\begin{equation}
\bar{\sigma}=\sqrt{17}-4\approx .1231...,
\label{4.x}
\end{equation}
as follows from the formula

\begin{equation}
\bar{\sigma}=\frac{1}{2}\sqrt{9\sigma^2+54\sigma+49}-\frac{3}{2}\sigma-\frac{7}{2}.
\label{4.x1}
\end{equation}
Since this is the only case relevant to the discussion below, we will from here on let 
$\bar{\sigma}$ denote the special value $\bar{\sigma}\equiv \sqrt{17}-4.$

\section{The Equation of State}\label{Sect4} 
\setcounter{equation}{0}

In this section we derive the FRW equation of state that
we use to close the system (\ref{4.33}) and (\ref{4.34}).
In this paper we consider the case when the equation of state for the 
FRW metric agrees with the equation of
state in the standard model of cosmology after the time that the 
radiation in the universe uncoupled from thermal equilibrium with matter.
This is an effort to account for the observed microwave 
backgound radiation level in our shock--wave model.  Our idea is that
if the expanding universe arose from a great explosion, then one might
conjecture that the expansion would have settled down to a uniform 
expansion by the time that this decoupling occured.  In the standard 
model of cosmology, the thermal uncoupling of radiation and matter
occured at about $300,000$ years after the {\it Big Bang}, at a
temperature of about $4000$ degrees Kelvin, \cite{blaugu,wein,peeb}.  Thus, 
we analyze our shock--wave model in the case that there is an energy
density $\rho_r$ for the radiation, (which supplies a pressure $p_r=(1/3)\rho_r$
via the Stefan-Boltzmann law),  and
a separate energy density $\rho_m$ for the matter, which is assumed to 
exert a
zero pressure $p_m=0.$ Since the scale factor $R$ is the 
independent variable in our shock equations (\ref{4.33}) and 
(\ref{4.34}), we now obtain formulas for the FRW energy density and
pressure as functions of $R.$

We start with the FRW equations in the form, \cite{wein},

\begin{equation}
\label{5.1}
\dot{R}^2=\frac{8\pi{\cal G}}{3c^4}\rho R^2-k,
\end{equation}
and

\begin{equation}
\label{5.2}
\frac{d}{dR}(\rho R^3)=-3pR^2.
\end{equation}
(Again, we assume that \lq\lq dot" denotes $d/d(ct).$)  We can rewrite
(\ref{5.1}) as
  
\begin{equation}
\label{5.3}
\dot{R}^2=QR^2-k,
\end{equation}
and  equation (\ref{5.2}) as

\begin{equation}
\label{5.4}
\frac{d}{dR}(QR^3)=-3PR^2,
\end{equation}
where $Q=\frac{8\pi{\cal G}}{3c^4}\rho$ and 
$P=\frac{8\pi{\cal G}}{3c^4}p$ have dimensions of inverse length
squared, c.f. (\ref{4.19}).  Now assume that the energy in the FRW system
is in the form of pure radiation and matter alone, so that   

\begin{equation}
\label{5.5a}
Q=Q_r+Q_m
\end{equation}
where $Q_r,$ $Q_m$ denote the (appropriatly scaled) energy density of
radiation and matter, respectively.  Further, assume that the pressure
of radiation is given by the Stefan-Boltzmann Law, \cite{wein},

\begin{equation}
\label{5.5}
P_r=(1/3)Q_r,
\end{equation}
and that

\begin{equation}
\label{5.6}
P_m=0,
\end{equation}
so that the uncoupling implies that

\begin{equation}
\label{5.7}
P=P_r+P_m=P_r=(1/3)Q_r.
\end{equation}
Finally, assume that 

\begin{equation}
\label{5.8}
Q_m=\frac{\beta}{R^3},
\end{equation}
for some positive constant $\beta,$ so that the total energy of matter
within a (geodesically) expanding volume in the FRW metric remains constant.  Substituting (\ref{5.5})
through (\ref{5.8}) into (\ref{5.4}) gives

\begin{equation}
\label{5.9}
\frac{\frac{d}{R}(Q_rR^3)}{Q_rR^3}=-\frac{1}{R},
\end{equation}
which has the solution

\begin{equation}
\label{5.10}
Q_r=\frac{3\alpha}{R^4},
\end{equation}
for some positive constant $\alpha.$ Then the FRW pressure is given by

\begin{equation}
\label{5.11}
P=P_r=\frac{\alpha}{R^4}.
\end{equation}
We conclude that the equation of state that applies to the FRW system
under the assumption that radiation is uncoupled from matter is given
by

\begin{equation}
\label{5.12}
Q=\frac{3\alpha}{R^4}+\frac{\beta}{R^3},
\end{equation}

\begin{equation}
\label{5.13}
P=\frac{\alpha}{R^4}.
\end{equation}
Putting (\ref{5.12}), (\ref{5.13}) into equations (\ref{4.39}),
(\ref{4.40}) gives the system of ODE's 

\begin{eqnarray}
\frac{dr}{dR}&=&\frac{R}{(3\alpha+\beta R-kR^2)}
\left(\frac{\alpha-\bar{P}R^4}{3\alpha+\beta R+\bar{P}R^4}\right)
\frac{(1-kr^2)}{r},\label{5.14}\\\nonumber\\
\frac{d\bar{P}}{dR}&=&-\frac{1}{2R^5}\frac{(3\alpha+\beta R+3\bar{P}R^4)(\alpha-\bar{P}R^4)}
{3\alpha+\beta R-kR^2},\label{5.15}
\end{eqnarray}
Observe, again, that a nice feature of the formulation (\ref{5.14}), (\ref{5.15}) is that, in this formulation,
the second equation (\ref{5.15}) for the TOV pressure $\bar{P}$ uncouples from the first 
equation (\ref{5.14}) for the shock position $r.$

Equations (\ref{5.12}) and (\ref{5.13}) together with the Stefan-Boltzmann law imply that the
temperature of radiation  is proportional to $1/R.$
Indeed, let $T\equiv T(R)$ denote the temperature of 
radiation.  The Stefan-Bolzmann law relates the energy density of radiation
$\rho_r$ to the temperature through the relation

\begin{equation}
\label{SB}
\rho_r=aT^4,
\end{equation}
where

\begin{equation}
\label{SB1}
a\approx7.664\times10^{-15}\frac{erg}{cm^3}(K^o)^4.
\end{equation}
\vspace{.3cm}
Since $Q_r=\frac{8\pi{\cal G}}{3c^4}\rho_r,$ we can write this as

\begin{equation}
\label{SB2}
Q_r=\hat{a}T^4,
\end{equation} 
where, \cite{wein},

\begin{equation}
\label{SB3}
\hat{a}=\frac{8\pi{\cal G}a}{3c^4}
\end{equation}
defines the constant $\hat{a}.$  Now  by (\ref{5.10})

$$
\frac{3\alpha}{R^4}=Q_r=\frac{8\pi{\cal G}}{3c^4}\rho_r=
\frac{8\pi{\cal G}}{3c^4}aT^4=\hat{a}T^4,
$$
we have the following lemma:

\begin{Lemma}
\label{L8.1}
The Stefan-Boltzmann law implies that

\begin{equation}
\label{5.16}
T=\left(\frac{3\alpha}{\hat{a}}\right)^{1/4}\frac{1}{R}.
\end{equation}
\end{Lemma}

\section{Restriction to $k=0$---Phase Plane Analysis}
\label{Sect5} 
\setcounter{equation}{0}

We now analyze system (\ref{5.14}), (\ref{5.15}) in the case of critical expansion when $k=0.$
The case $k\neq0$ will be considered in a subsequent paper.   To start, recall that in the case $k=0,$ system 
(\ref{4.39}), (\ref{4.40}) reduces to the (non-autonomous) system

\begin{eqnarray}
\frac{dr}{dR}&=&\frac{1}{(QR^3)}\left(\frac{P-\bar{P}}
{Q+\bar{P}}\right)\left(\frac{1}{r}\right),\label{6.1}\\\nonumber\\
\frac{d\bar{P}}{dR}&=&-\frac{1}{2}\frac{(Q+3\bar{P})(P-\bar{P})}
{QR},\label{6.2}
\end{eqnarray} 
where again 

\begin{equation}
\label{6.2a}
(Q,P,\bar{Q},\bar{P})=\frac{8\pi{\cal G}}{3c^4}(\rho,p,\bar{\rho},\bar{p}),
\end{equation}
all have the dimensions of inverse length squared.  Assuming now that $Q$ and $P$ are given by
(\ref{5.12}), (\ref{5.13}), 
$$
Q=\frac{3\alpha}{R^4}+\frac{\beta}{R^3},
$$
$$
P=\frac{\alpha}{R^4},
$$
(which models a universe of matter and radiation assuming no thermal coupling),
and substituting this into (\ref{6.1}), (\ref{6.2}) yields the system

\begin{eqnarray}
\frac{dr}{dR}&=&\frac{R}{(3\alpha+\beta R)}
\left(\frac{\alpha-\bar{P}R^4}{3\alpha+\beta R+\bar{P}R^4}\right)
\frac{1}{r},\label{6.3}\\\nonumber\\
\frac{d\bar{P}}{dR}&=&-\frac{1}{2R^5}\frac{(3\alpha+\beta R+3\bar{P}R^4)(\alpha-\bar{P}R^4)}
{3\alpha+\beta R},\label{6.4}
\end{eqnarray}
which is just system (\ref{5.14}), (\ref{5.15}) in the case $k=0.$  Solutions of 
system (\ref{6.3}) and (\ref{6.4}) determine the shock position $r(R),$ (the position as measured by
the radial coordinate of the FRW metric that is behind the shock--wave),
together with the TOV pressure $\bar{P}(R)$
in front of the shock.  The TOV pressure $\bar{p}(\bar{r})$ is then recovered from the
solution $(r(R),\bar{P}(R))$ by inverting the equation $\bar{r}=Rr(R)$ and using 
$\bar{P}=\frac{8\pi{\cal G}}{3c^4}\bar{p}.$  The function $R(t)$ is 
obtained by solving the 
FRW equation (\ref{4.1}) with $k=0,$ and due to the scaling law for this equation, we are free to choose 
the scale factor $R_0$ such that $R_0=1$ at present time in the universe.  The constants 
$\alpha$ and $\beta$ that determine $Q$ rescale with choice of $R_0,$
and are determined from initial conditions for the FRW metric.  For an FRW metric that models
the expanding universe, we can take one of the two initial conditions as

\begin{equation}
\label{6.3a}
Q_0=3\alpha+\beta=H_0^2,
\end{equation}
where $H_0$ is the present value of the Hubble constant, c.f. (\cite{peeb}).  
Finally, the TOV energy density 
$\bar{Q}$ is given by the formula in (\ref{fill4}): 

\begin{equation}
\label{6.3b}
\bar{Q}=\frac{1}{3\bar{r}^2}\frac{d}{d\bar{r}}(Q\bar{r}^3)=Q+\frac{Rr}{3}\frac{d}{d\bar{r}}Q.
\end{equation}
This simplifies under our special assumption (\ref{5.12}) for $Q.$  Indeed, by (\ref{5.12}),

\begin{equation}
\label{6.3c}
\frac{dQ}{d\bar{r}}=\frac{dQ}{dR}\frac{dR}{d\bar{r}}=
\frac{d}{dR}\left(\frac{3\alpha}{R^4}+\frac{\beta}{R^3}\right)
\left(r+R\frac{dr}{dR}\right)^{-1}.
\end{equation}
Since

\begin{equation}
\label{6.3d}
\frac{d\bar{r}}{dR}=\frac{d(Rr)}{dR}=r+R\frac{dr}{dR},
\end{equation}
where $\frac{dr}{dR}$ is given by equation (\ref{6.1}).  
Putting (\ref{6.1}) and (\ref{6.3c}) into (\ref{6.3b}) and
simplifying yields the following expression for the TOV energy density $\bar{Q}:$

\begin{equation}
\label{6.4c}
\bar{Q}=Q-\frac{(4\alpha+\beta R)(3+\frac{\beta}{\alpha}R+w)(3+\frac{\beta}{\alpha}R)\alpha r^2}
{R^4\left\{\alpha r^2(3+\frac{\beta}{\alpha}R+w)
(3+\frac{\beta}{\alpha}R)+(1-w)R^2\right\}},
\end{equation}
where

\begin{equation}
w=\frac{\bar{P}R^4}{\alpha}.
\label{6.5a}
\end{equation}
We conclude that each choice
of constants $\alpha$ and $\beta$ and each choice of
initial conditions for (\ref{6.3}) and (\ref{6.4}) determines a shock--wave solution of the 
Einstein equations, at each point where all the variables are positive.  
We now analyze solutions 
of system (\ref{6.3}) and (\ref{6.4}) in detail. 

Substituting $w$ for $\bar{P}$ in (\ref{6.3}), (\ref{6.4}) yields the equivalent
system

\begin{eqnarray}
\frac{dr}{dR}&=&\frac{R}{(3+\frac{\beta}{\alpha} R)}
\left(\frac{1-w}{3+\frac{\beta}{\alpha} R+w}\right)
\frac{1}{\alpha r},\label{6.6}\\\nonumber\\
\frac{dw}{dR}&=&\frac{4w}{R}
\left(1-\frac{(1-w)(3+\frac{\beta}{\alpha} R+3w)}
{8(3+\frac{\beta}{\alpha} R)w}\right).\label{6.7}
\end{eqnarray}
Now (\ref{6.7}) is a non-autonomous scalar equation for $\bar{P}$ that uncouples from equation (\ref{6.6}).
In order to do a phase plane analysis of equation (\ref{6.7}), and to analyze the behavior of solutions as
$R\rightarrow\infty,$ we now rewrite (\ref{6.7}) as an autonomous system.  To this end, set 

\begin{equation}
S=\frac{1}{\alpha}R,
\label{6.5aa}
\end{equation}
and 

\begin{equation}
u=\frac{1}{S}.
\label{6.5c}
\end{equation}
Substituting these into (\ref{6.7}) and letting \lq\lq dot" 
denote $d/dS,$ we obtain the following autonomous system of two ODE's that
is equivalent to equation (\ref{6.7}):

\begin{eqnarray}
\dot{u}\equiv\frac{du}{dS}&=&-u^2,\label{6.8}\\\nonumber\\
\dot{w}\equiv\frac{dw}{dS}&=&4wu\left[1-\frac{(1-w)(\beta+3(1+w)u)}
{8(\beta+3u)w}\right].\label{6.9}
\end{eqnarray}
We now analyze the phase plane associated with system (\ref{6.8}), (\ref{6.9}).

System (\ref{6.8}), (\ref{6.9}) has a line of rest points at $u=0,$ and an isocline 
where $\dot{w}=0.$   Setting the RHS of equation (\ref{6.9}) equal to zero gives

\begin{equation}
\left[1-\frac{(1-w)(\beta+3(1+w)u)}
{8(\beta+3u)w}\right]=0,
\label{6.10}
\end{equation}
and solving this for $w$ gives
 
\begin{equation}
w=\frac{(3\beta+8u)}{2u}\left\{-1+\sqrt{1+\frac{4}{3}
\frac{(\beta+3u)}{(3\beta+8u)^2}u}\right\}\equiv\phi(u).
\label{6.11}
\end{equation}
Thus the isocline is defined for $0<u<\infty$ by

\begin{equation}
w=\phi(u).
\label{6.12}
\end{equation}
Note first that when $\beta=0,$ (the case of pure radiation), the isocline degenerates to

\begin{equation}
\phi(u)\equiv\sqrt{17}-4=\bar{\sigma}.
\label{6.13}
\end{equation}
It is straightforward to verify that when $\beta=0,$ the isocline is also a solution orbit
of system (\ref{6.8}), (\ref{6.9}),  and the special solution
in \cite{smolte1}, discussed above starting with (\ref{4.a})-(\ref{4.d}), corresponds to this 
orbit. The special value $\bar{\sigma}=\sqrt{17}-4$ also is important in the case $\beta\neq0.$ 
The next theorem gives the qualitative behavior of the solution
orbits of system (\ref{6.8}), (\ref{6.9}) in the $(u,w)$-plane when $\beta\neq0.$

\begin{Theorem}
\label{theorem6.1}  
Assume that $\beta\neq0.$  Then the following statements are true 
regarding the
phase plane of system (\ref{6.8}), (\ref{6.9})).  
(Here we define an orbit of system 
(\ref{6.8}), (\ref{6.9}) to be a function $w=w(u)$ such that $(u(S),w(1/S)),$ ($u=1/S$), 
is a solution of system (\ref{6.8}), (\ref{6.9})): 
\vspace{.2cm}

\noindent (i) The isocline $w=\phi(u)$ defined by $(\ref{6.11})$ is monotone increasing
for $0<u<\infty,$ and satisfies
\begin{eqnarray}
\lim_{u\rightarrow\infty}\phi(u)&=&\bar{\sigma}\equiv\sqrt{17}-4\approx .1231...,
\label{6.13.i}\\
\lim_{u\rightarrow0}\phi(u)&=&\frac{1}{9}\approx.1111...,\label{6.13.ii}\\
\lim_{u\rightarrow0}\phi'(u)&=&\frac{1-\left(\frac{1}{9}\right)^2-\frac{8}{9}}{9\beta}
\approx \frac{.01097...}{\beta}>0.\label{6.13.iii}
\end{eqnarray}
\vspace{.2cm}

\noindent (ii)  Orbits can only cross the isocline $w=\phi(u)$ once, from right to 
left in the $(u,w)$-plane, as $S$ increases, (see Fig 1.).
\vspace{.2cm}

\noindent (iii)  Along any orbit $w=w(u)$ we have
\begin{equation}
\label{6.14}
\lim_{u\rightarrow\infty}w(u)=\bar{\sigma}.
\end{equation}
\vspace{.2cm}

\noindent (iv)  There exists a unique orbit $w_{crit}(u)$ satisfying
\begin{equation}
\label{6.15}
\lim_{u\rightarrow0}w_{crit}(u)=\frac{1}{9}.
\end{equation}
Moreover, all orbits $w=w(u)$ starting from initial conditions $(u_0,w_0)$ such that
$w_0>\phi(u_0),$ (that is, starting above the isocline), satisfy
\begin{equation}
\label{6.16}
\lim_{u\rightarrow0}w(u)=\infty;
\end{equation}
and all orbits starting from initial conditions $(u_0,w_0)$ such that
$w_0<\phi(u_0),$ (starting below the isocline), satisfy
\begin{equation}
\label{6.17}
\lim_{u\rightarrow0}w(u)=-\infty.
\end{equation}

\end{Theorem}
\vspace{.2cm}

\noindent {\bf Proof:}  To verify (\ref{6.13.i}), we have

\begin{eqnarray}
\lim_{u\rightarrow\infty}\phi(u)&=&
\lim_{u\rightarrow\infty}\frac{(3\beta+8u)}{2u}\left\{-1+\sqrt{1+\frac{4}{3}
\frac{(\beta+3u)}{(3\beta+8u)^2}u}\right\}\nonumber\\
&=&\lim_{u\rightarrow\infty}\frac{(3\beta+8u)}{2u}
\left\{-1+\sqrt{1+\frac{1}{16}}\right\}\nonumber\\
&=&-1+\sqrt{17}\equiv\bar{\sigma}.\nonumber
\end{eqnarray}

We next show that $\phi(u)$ tends to $\bar{\sigma}$ monotonically from below as
$u\rightarrow\infty.$  Note that by (\ref{6.11}), $w=\phi(u)$ is equivalent to

\begin{equation}
0=\left[1-\frac{(1-w)(\beta+3(1+w)u)}
{8(\beta+3u)w}\right],
\nonumber
\end{equation}
which we rewrite as
\begin{equation}
0=8(\beta+3u)w-(1-w)\left[\beta+3(1+w)u\right].
\label{6.18}
\end{equation}

\noindent Now differentiating (\ref{6.18}) implicitly with respect to $u$ gives
\begin{eqnarray}
8(\beta+3u)\frac{dw}{du}+24w&=&-\left[\beta+3(1+w)u\right]\frac{dw}{du}\nonumber\\
\ \ &&+
(1-w)\left[3+3w+3u\frac{dw}{du}\right].
\label{6.19}
\end{eqnarray}
Simplifying (\ref{6.19}) we obtain

\begin{eqnarray}
(9\beta+24u+6uw)\frac{dw}{du}=-3(w^2+8w-1).
\label{6.20}
\end{eqnarray}
Now the roots of $w^2+8w-1$ are

\begin{equation}
\bar{\sigma}\equiv\sqrt{17}-4\approx.1231,\ \ \ \tilde{\sigma}\equiv-\sqrt{17}-4,
\label{6.21}
\end{equation}
and thus we conclude that, along the isocline $w=\phi(u),$ 

\begin{eqnarray}
\frac{dw}{du}>0\ \ {\rm if}\ \ w<\bar{\sigma},\\
\frac{dw}{du}<0\ \ {\rm if}\ \ w>\bar{\sigma},
\label{6.22}
\end{eqnarray}
where we use the fact that $\phi(u)>0$ for all $u.$  
Thus, it suffices
to show that $\phi(u)\neq\bar{\sigma}$ for any $u$ in order to conclude that $\phi'(u)\neq0$ for
$0<u<\infty.$  So assume for contradiction that $\phi(u)=\bar{\sigma}.$  But solving for
$u$ in (\ref{6.18}) gives

\begin{equation}
u=-\frac{\beta(9w-1)}{(w-\bar{\sigma})(w+|\tilde{\sigma}|)}\ ,
\label{6.23}
\end{equation}
and thus $w=\bar{\sigma}$ leads to a contradiction unless $\beta=0.$  We conclude that
if $\beta\neq0,$ then $\phi(u)$ monotonically increases to $\bar{\sigma}$ as
$u\rightarrow\infty,$ thus proving (\ref{6.13.i}).  Statement 
(\ref{6.13.ii}) follows from (\ref{6.18}), and (\ref{6.13.ii}) follows from 
(\ref{6.20}).  Thus the proof of (i) is
complete. 

Statement (ii) follows because $\dot{w}=0$ only on $w=\phi(u),$ 
$\dot{w}>0$ if $w>\phi(u),$ and since we have shown that
$\phi'(u)>0,$ it follows that orbits can only cross the isocline from right to left in
forward $S$-time.

To verify (iii), we show that all orbits tend in backward time, (increasing $u$), to 
$w=\bar{\sigma}.$  To see this note that

\begin{eqnarray}
\lim_{u\rightarrow\infty}\dot{w}&=&\lim_{u\rightarrow\infty}4wu
\left\{1-\frac{(1-w)(3+\frac{\beta}{u}+3w)}{8(3+\frac{\beta}{u}w)}\right\}\nonumber\\
&\approx&4wu\left\{1-\frac{1-w^2}{8w}\right\}
\label{6.24}
\end{eqnarray}

\noindent where approximately means to leading order as 
$u\rightarrow\infty.$  Now each orbit that starts above
$w=\phi(u)$ decreases as u increases unless the orbit crosses the isocline, in which case
the orbit increases from there on out as $u\rightarrow\infty.$  It follows that orbits starting
below $w=\phi(u)$ can never cross $w=\phi(u)$ at any value of $u$ larger than the initial
value.  Thus, since $\lim_{u\rightarrow\infty}\phi(u)=\bar{\sigma},$ all orbits must be bounded
above in $w$ by the maximum of $\left\{\bar{\sigma},w_0\right\},$ and bounded below by the
minimum of $\left\{1/9,w_0\right\}.$  But from (\ref{6.24}), we must have that

\begin{equation}
\label{6.25}
\lim_{u\rightarrow\infty}\left\{1-\frac{1-w^2}{8w}\right\}=0.
\end{equation}
Indeed, if not, then (\ref{6.24}) implies that $\left|\dot{w}\right|$ tends to infinity as $u\rightarrow\infty,$ which
implies that $w$ is not bounded as $u\rightarrow\infty,$ and this contradicts the above bounds.
Since $\bar{\sigma}$ is the only positive root of $\left\{1-\frac{1-w^2}{8w}\right\},$ we conclude
from (\ref{6.25}) that

$$
\lim_{u\rightarrow\infty}w=\bar{\sigma}.
$$
This completes the proof of (iii).

We now give the proof of (iv).  Our approach for this is to write the ODE for $w$ as a function
of $u$ along orbits, and study the limit $u\rightarrow0.$  From 
(\ref{6.8}) and (\ref{6.9}),

\begin{eqnarray}
-\frac{dw}{du}&=&\frac{4w}{u}\left[1-\frac{(1-
w)(\beta+3(1+w)u)}{8(\beta+3u)w}\right].\nonumber\\
&\approx&\frac{4w}{u}\left[1-\frac{1-w}{8w}\right],
\label{6.26}
\end{eqnarray}
where approximate equality means to leading order as $u\rightarrow0.$  Now assume for 
contradiction that there exists an orbit $w=f(u)$ that is bounded in a neighborhood of
$u=0,$ but such that $\lim_{u\rightarrow0}f(u)\neq1/9.$  The boundedness condition implies that
(\ref{6.26}) applies with errors that are bounded as $u\rightarrow0.$  That is,

\begin{eqnarray}
-\frac{dw}{du}&=&\frac{4w}{u}\left(1-\frac{1-w}{8w}\right)+O(1)\noindent\\
&=&\frac{9w-1}{2u}+O(1),
\label{6.27}
\end{eqnarray}
where $O(1)$ denotes a constant that depends on the bounds for $w$ but is independent of $u$ as $u\rightarrow0.$
Integrating (\ref{6.27}) leads to the estimate

\begin{eqnarray}
-\frac{9w-1}{9w_0-1}&=&\left(\frac{u_0}{u}\right)^{9/2}+O(1)e^{O(1)|u-u_0|},
\label{6.28}
\end{eqnarray}
where $(u_0,w_0)$ are taken as initial data, $u_0>0.$  But (\ref{6.28}) implies that if
$\lim_{u\rightarrow0}w\neq1/9,$ then $w=f(u)$ is unbounded near $u=0.$  From this we conclude
that every orbit that is bounded as $u\rightarrow0$ satisfies

\begin{eqnarray}
\lim_{u\rightarrow0}w=1/9. 
\label{6.29}
\end{eqnarray}

We now show that there exists at least one orbit such that $\lim_{u\rightarrow0}w=1/9.$  Note 
first that any orbit starting from initial data $(u_0,w_0)$  that lies on the isocline, 
$w_0=\phi(u_0),$ $u_0>0,$ must lie above the isocline for all $0<u<u_0$ because 
we know that $\frac{dw}{du}<0$ on this interval, and $\phi'(u)>0.$    Since the isocline 
decreases to $\bar{\sigma}\approx.1231>1/9$ as $u\rightarrow0,$  it follows that $w_0>1/9$ for 
initial data lying above the isocline, and hence $\lim_{u\rightarrow0}w>1/9$ along an orbit starting
from such initial data.
But our argument above shows that when this happens, we must have $\lim_{u\rightarrow0}w=+\infty.$  
We conclude that $\lim_{u\rightarrow0}w=+\infty$ for any orbit
starting from initial data above the isocline, $w_0>\phi(u_0).$  Similarly, if the initial data 
$(u_0,w_0)$ lies below the line $w=1/9,$ that is, $w_0<1/9,$ then also $w_0<\phi(u_0)$ because we have
that $\phi(u)>1/9.$   Thus from (\ref{6.26}), $\frac{dw}{du}>0,$ and so it follows that $\lim_{u\rightarrow0}<1/9,$
and our argument above implies that $\lim_{u\rightarrow0}w=-\infty.$  We conclude that $\lim_{u\rightarrow0}w=-\infty$ for any 
orbit
starting from initial data below the line $w=1/9;$ and 
$\lim_{u\rightarrow0}w=+\infty$ for any orbit
starting from initial data above the isocline, $w_0>\phi(u_0).$   Now consider all orbits emanating from 
initial data on some fixed vertical line $u=\epsilon>0.$  Then if $w_0>\phi(\epsilon),$ we have 
$\lim_{u\rightarrow0}=+\infty;$ and if $w_0<1/9,$ we have $\lim_{u\rightarrow0}=-\infty.$  So define

\begin{equation}
\label{6.30}
w_{+}=Inf\left\{w_0:\lim_{u\rightarrow0}w=+\infty\right\},
\end{equation}   
where the limit is taken along the orbit emanating from the point $(\epsilon,w_0).$  
We now claim that the critical orbit emanating from initial condition $(\epsilon,w_0)$ satisfies 
$\lim_{u\rightarrow0}w=1/9.$  To see this note first that $w_+\geq1/9$ because orbits below
$w=1/9$ tend to $-\infty$ as $u\rightarrow0.$  We show next that the orbit emanating from $(\epsilon,w_+)$ cannot tend to $w=+\infty$
as $u\rightarrow0.$  To see this, note that if $\lim_{u\rightarrow0}w=+\infty$ along the crititcal orbit,
then this must be true for all orbits starting in a neighborhood of $(\epsilon,w_0)$ as well.  Indeed, if 
$\lim_{u\rightarrow0}w=+\infty,$ then at some positive value of $u$ we must have $w>\phi(u)$
along the critical orbit; and so by continuity, nearby orbits must also rise above the isocline
at some $u>0,$ and hence by above we know that $\lim_{u\rightarrow0}w=+\infty$ along 
orbits sufficiently close to the critical orbit.  But this contradicts the fact that $w_-$ is a 
greatest lower bound.  We conclude that we cannot have $\lim_{u\rightarrow0}w=+\infty$ along the 
critical orbit.  Similarly, we cannot have $\lim_{u\rightarrow0}w=-\infty$ along the critical orbit
because then nearby orbits would also satisfy $\lim_{u\rightarrow0}w=-\infty$ since 
they would cross $w=1/9$ before $u=0,$ and again this would contradict the fact that $w_+$ is a
greatest lower bound.  Since we cannot have $\lim_{u\rightarrow0}w=-\infty$ or
$\lim_{u\rightarrow0}w=+\infty,$ it follows from (\ref{6.29}) that the only alternative 
is that $\lim_{u\rightarrow0}w=1/9$ along the critical orbit, as claimed.

We now show that the critical orbit is unique.  To this end, rewrite equation (\ref{6.9}) as

\begin{equation}
\label{6.31}
\dot{w}=4wu\left[1-\frac{(1-w)(\beta+3(1+w)u)}
{8(\beta+3u)w}\right]\equiv F(u,w).
\end{equation}
Differentiating (\ref{6.31}) with respect to $w$ gives

\begin{equation}
\label{6.32}
\frac{d\dot{w}}{dw}\equiv\frac{\partial F}{\partial w}=
4u\left[1+\frac{6w+\beta/u}{8\left(3+\beta/u\right)}\right]>0.
\end{equation}
But (\ref{6.32}) implies that the distance between orbits is increasing in forward time $S,$ (that is,
increasing as $u=1/S$ decreases.  Indeed,

\begin{equation}
\label{6.33}
\dot{(w_2-w_1)}=F(u,w_2)-F(u,w_1)=\frac{\partial F}{\partial w}(u,w_*)(w_2-w_1)>0
\end{equation}
if $w_2-w_1>0.$  This implies that there cannot be two orbits that satisfy
$\lim_{u\rightarrow0}w=1/9$ since the distance between them would then tend to zero as
$u\rightarrow0,$ contradicting (\ref{6.33}).  This finishes the proof of (iv), and thus the 
proof of the theorem is complete.
\vspace{.2cm}

The salient properties of the phase plane 
for system (\ref{6.8}), (\ref{6.9}) are sketched in Figure 1.  Note that as $\beta\rightarrow0,$
the isocline moves up to the line $w=\bar{\sigma},$ (continuously, except for a jump from $1/9$
to $\bar{\sigma}$ at $u=0,$ $\beta=0$).  The isocline is a curve of absolute minima of orbits that
cross the isocline, and the isocline, together with all orbits, tend to $w=\bar{\sigma}$ as
$u\rightarrow\infty,$ $R\rightarrow\infty.$  Moreover, all orbits except the critical orbit 
tend to infinity as
$u\rightarrow0,$ ($R\rightarrow\infty$), and so the critical orbit is the only orbit bounded for all
values of $R>0.$  Along both the critical orbit and the isocline, the following apriori bounds hold for all
$0<R<\infty:$
 
\begin{equation}
\label{7.1a}
1/9\approx .1111<w<\bar{\sigma}\approx .1231.
\end{equation} 
Note, however, that the critical orbit and the isocline do not coincide except in the limiting
case $\beta=0,$ in which case both reduce to the line $w=\bar{\sigma},$  which also can be indentified
with the special solution constructed in \cite{smolte1}.  In particular, Figure 1 describes
how this special solution is imbedded in the larger class of solutions that allow for general initial data.

\section{Conditions for $\bar{Q}>0$ and $\bar{Q}>\bar{P}$}
\label{Sect6} 
\setcounter{equation}{0}

In this section we obtain conditions which guarantee that $\bar{Q}>0$ and $\bar{Q}>\bar{P},$ physically
reasonable conditions on the TOV energy density and pressure, ($\bar{Q}\equiv\frac{8\pi{\cal G}}{c^4}\bar{\rho}$ where
$\bar{\rho}$ is true TOV energy density.)  In particular, $\bar{Q}>\bar{P}$ guarantees that
$\bar{Q}>0$ whenever the solution orbits of system (\ref{6.8}), (\ref{6.9}) satisfy $w\equiv\bar{P}R^4/\alpha>0.$
We begin with the formula (\ref{6.4c}) for $\bar{Q}:$

\begin{equation}
\bar{Q}=Q-\frac{(4\alpha+\beta R)(3+\frac{\beta}{\alpha}R+w)(3+\frac{\beta}{\alpha}R)\alpha r^2}
{R^4\left\{\alpha r^2(3+\frac{\beta}{\alpha}R+w)(3+\frac{\beta}{\alpha}R)+(1-w)R^2\right\}}.\nonumber
\end{equation}
Asking that the RHS of (\ref{6.4c}) be positive, and using the formula

$$
Q=\frac{3\alpha}{R^4}+\frac{\beta}{R^3}
$$
we see that $\bar{Q}>0$ is equivalent to

\begin{eqnarray}
\alpha\left\{\alpha r^2(3+\frac{\beta}{\alpha}R+w)(3+\frac{\beta}{\alpha}R)+(1-
w)R^2\right\}\label{7.1}\\
-(4\alpha+\beta R)(3+\frac{\beta}{\alpha}R+w)\alpha r^2>0.\nonumber
\end{eqnarray}
Solving (\ref{7.1}) for $r^2$ leads to the following inequality that is equivalent to
$\bar{Q}>0:$

\begin{eqnarray}
\alpha r^2<\frac{(1-w)R^2}{3+\frac{\beta}{\alpha}R+w}.\label{7.2}
\end{eqnarray}
Equation (\ref{7.2}) implies that the condition $\bar{Q}>0$ puts a contraint on the maximum
possible shock position at a given value of $R.$  The following theorem implies that if the
condition holds at some value $R=R_*$ in a solution of (\ref{6.8}), (\ref{6.9}), then it holds for all
$R\geq R_*$ in that solution, so long as $0<w<1$ and $dw/dR<0.$  Both of these conditions are satisfied
along the critical orbit where $1/9<w<\bar{\sigma}\approx.1231.$ 

\begin{Lemma}
\label{T7.1}
Define the quantity $\{\}_I$ by

\begin{equation}
\label{7.4}
\{\}_I\equiv\left\{\frac{(1-w)R^2}{3+\frac{\beta}{\alpha}R+w}-\alpha r^2\right\}_I.
\end{equation}
Then for any solution of (\ref{6.6}), (\ref{6.7}) we have

\begin{eqnarray}
\frac{d}{dR}\{\}_I>0 \label{7.5}
\end{eqnarray}
at each point where

\begin{eqnarray}
0<w<\bar{\sigma}, \label{7.3a}
\end{eqnarray}
and

\begin{eqnarray}
\frac{dw}{dR}<0. \label{7.3}
\end{eqnarray}

\end{Lemma}
Lemma \ref{T7.1} implies that $\{\}_I$ is monotone increasing along any solution of (\ref{6.8}), (\ref{6.9}) that
satisfies (\ref{7.3}), and thus if (\ref{7.2}) holds at a value $R=R_*$ in such a solution, then it
must hold at all $R>R_*.$
\vspace{.3cm}

\noindent{\bf Proof:} Starting with (\ref{7.4}) we have

\begin{eqnarray}
\label{7.6}
\frac{d}{dR}\{\}_I = \frac{2(1-w)R}{3+\frac{\beta}{\alpha}R+w}
-\alpha\frac{dr^2}{dR}+R^2\frac{d}{dR}\left\{\frac{1-w}{3+\frac{\beta}{\alpha}R+w}\right\}_{II}\\
=\frac{2(1-w)R}{3+\frac{\beta}{\alpha}R+w}
-\frac{2(1-w)R}{(3+\frac{\beta}{\alpha}R+w)(3+\frac{\beta}{\alpha}R)}
+R^2\frac{d}{dR}\{\}_{II}\nonumber
\end{eqnarray}
where we have used (\ref{6.6}).  This simplifies to

\begin{eqnarray}
\label{7.7}
\frac{d}{dR}\{\}_I=\frac{2(1-w)R}{3+\frac{\beta}{\alpha}R+w}\left(
\frac{2+\frac{\beta}{\alpha}R}{3+\frac{\beta}{\alpha}R}\right)+R^2\frac{d}{dR}\{\}_{II}.
\end{eqnarray}
Moreover,

\begin{eqnarray}
\nonumber
\frac{d}{dR}\{\}_{II}&=&\frac{d}{dR}\left\{\frac{1-
w}{3+\frac{\beta}{\alpha}R+w}\right\}_{II}\\\nonumber
&=&\frac{(3+\frac{\beta}{\alpha}R+w)\left(-\frac{dw}{dR}\right)-(1-w)(\frac{\beta}{\alpha}+\frac{dw}{dR})}
{(3+\frac{\beta}{\alpha}R+w)^2}\\\nonumber
&=&\frac{(3+\frac{\beta}{\alpha}R+1)\left(-\frac{dw}{dR}\right)-(1-w)\frac{\beta}{\alpha}}
{\left(3+\frac{\beta}{\alpha}R+w\right)^2}\\\label{7.8}
&\geq&-\frac{(1-w)\frac{\beta}{\alpha}}{\left(3+\frac{\beta}{\alpha}R+w\right)^2}>0.
\end{eqnarray}
where we have used (\ref{7.3a}) and (\ref{7.3}).  Using (\ref{7.8}) in (\ref{7.7}) and
simplifying gives

\begin{eqnarray}
\label{7.9}
\frac{d}{dR}\{\}_I\geq\frac{(1-w)\left(4+\frac{\beta}{\alpha}R\right)R}{\left(3+\frac{\beta}{\alpha}R+w)\right)
\left(3+\frac{\beta}{\alpha}R\right)}>0.
\end{eqnarray}
This completes the proof of Lemma \ref{T7.1}.  

We now obtain a corresponding condition for
$\bar{Q}>\bar{P}.$  Using (\ref{6.5a}) and (\ref{7.3a}) we know

\begin{equation}
\label{7.10}
\bar{P}\leq\frac{\alpha\bar{\sigma}}{R^4}.
\end{equation}
Using this together with the formula (\ref{6.4c}) for $\bar{Q},$ it
follows that the condition
$\bar{Q}>\bar{P}$ will hold if

\begin{equation}
\label{7.11}
\frac{R^4\bar{Q}}{\alpha}=\frac{3+\frac{\alpha}{\beta}R}
{\alpha r^2(3+\frac{\beta}{\alpha}R)+\frac{R^2(1-w)}{3+\frac{\beta}{\alpha}R+w}}
\{\}_I\geq\bar{\sigma}.
\end{equation}
Solving (\ref{7.11}) for $\alpha r^2$ gives the equivalent condition

\begin{equation}
\label{7.12}
\frac{1+\bar{\sigma}}{1-\frac{\bar{\sigma}}{3+\frac{\beta}{\alpha}R}}\alpha r^2\leq
\{\}_{III},
\end{equation}
where

\begin{equation}
\label{7.13}
\{\}_{III}=\frac{R^2(1-w)}{3+\frac{\beta}{\alpha}R+w}.
\end{equation}
Thus to get (\ref{7.12}) it suffices to have 

\begin{equation}
\label{7.14}
\frac{1+\bar{\sigma}}{1-\frac{\bar{\sigma}}{3}}\alpha r^2=(1+\epsilon)\alpha r^2\leq
\{\}_{III},
\end{equation}
where

\begin{equation}
\label{7.15}
\epsilon=\frac{4\bar{\sigma}}{3-\bar{\sigma}}.
\end{equation}
We conclude that $\bar{Q}\geq\bar{P}$ holds so long as

\begin{equation}
\label{7.16}
\left\{\{\}_{III}-(1+\epsilon)\alpha r^2\right\}\geq0.
\end{equation}

\begin{Lemma}
\label{T7.2}
Define the quantity $\{\}_{IV}$ by

\begin{equation}
\label{7.17}
\{\}_{IV}\equiv\left\{\{\}_{III}-(1+\epsilon)\alpha r^2\right\}.
\end{equation}
Then for any solution of (\ref{6.6}), (\ref{6.7}) we have

\begin{eqnarray}
\frac{d}{dR}\{\}_{IV}>0 \label{7.18}
\end{eqnarray}
at each point where (\ref{7.3a}) and (\ref{7.3}) hold.

\end{Lemma}

In particular, Lemma \ref{T7.2} implies that if (\ref{7.16}) holds at a point $R_*$ in a
solution of (\ref{6.6}), (\ref{6.7}) such that (\ref{7.3a}) and (\ref{7.3}) hold for all $R\geq R_*,$
then we conclude that (\ref{7.18}) holds at all points $R\geq R_*,$ and thus that $\bar{Q}\geq\bar{P}$ for
all $R\geq R_*.$
\vspace{.3cm}

\noindent{\bf Proof:}  Differentiating we obtain

\begin{eqnarray}
\frac{d}{dR}\{\}_{IV} &=& \frac{2(1-w)R}{3+\frac{\beta}{\alpha}R+w}
-(1+\epsilon)\alpha\frac{dr^2}{dR}+R^2\frac{d}{dR}\left\{\frac{1-w}{3+\frac{\beta}{\alpha}R+w}\right\}_{II}\nonumber\\
&=&\frac{2(1-w)R}{3+\frac{\beta}{\alpha}R+w}
-(1+\epsilon)\frac{2(1-w)R}{(3+\frac{\beta}{\alpha}R+w)(3+\frac{\beta}{\alpha}R)}
+R^2\frac{d}{dR}\{\}_{II}\nonumber\\
&=&\frac{2(1-w)R}{3+\frac{\beta}{\alpha}R+w}
\left(\frac{2-
\epsilon+\frac{\alpha}{\beta}R}{3+\frac{\alpha}{\beta}R}\right)+R^2\frac{d}{dR}\{\}_{II}.\label{7.19}
\end{eqnarray} 
Now using (\ref{7.8}) in (\ref{7.19}) and simplfying yields

\begin{eqnarray}
\frac{d}{dR}\{\}_{IV}\geq\frac{(1-w)R}{3+\frac{\beta}{\alpha}R+w}{3+\frac{\beta}{\alpha}R}
(4+\frac{\beta}{\alpha}R-2\epsilon)>0.\label{7.20}
\end{eqnarray} 
since $2\epsilon<4$ and $w<1.$  This concludes the proof of Lemma \ref{T7.2}. 
We have proven the following theorem:

\begin{Theorem}
\label{T7.3}
Assume that (\ref{7.3a}) and (\ref{7.3}) hold for all $R>R_*$ on a solution of (\ref{6.6}), (\ref{6.7}). 
Then $\{\}_I>0$ at $R=R_*$ is equivalent to $\bar{Q}>0$ at $R=R_*,$ and implies that $\bar{Q}>0$ for all $R>R_*;$ 
and if $\{\}_{IV}>0$ at $R=R_*$, then we must have $\bar{Q}>\bar{P}$ for all $R>R_*.$  The
condition $\{\}_I>0$ is equivalent to

\begin{equation}
\label{7.21}
\alpha r^2<\frac{(1-w)R^2}{3+\frac{\beta}{\alpha}R+w},
\end{equation}
and the condition $\{\}_{IV}>0$ simplifies to

\begin{equation}
\label{7.22}
\alpha r^2<\left(\frac{1-\bar{\sigma}/3}{1+\bar{\sigma}}\right)\frac{(1-w)R^2}{3+\frac{\beta}{\alpha}R+w}.
\end{equation}

\end{Theorem}

\section{Estimates for the Shock Position}
\label{Sect7} 
\setcounter{equation}{0}

In this section we take system (\ref{6.6}), (\ref{6.7}) as a simple cosmological model
in which the FRW metric behind the shock--wave at position $r$ is assumed to model the
expanding universe.  Given this, we now estimate the position of the shock--wave in the 
present universe as
determined by this model.  In this model, the expanding universe is modeled by an FRW, ($k=0,$)
metric in which the energy density $Q$ and pressure $P$ are given by (\ref{5.12}), (\ref{5.13}), that is, the same as that
assumed in the standard cosmological model after the time of thermal decoupling of matter with radiation, 
(approximately
300,000 years after the {\it Big Bang} in the standard model, \cite{wein}). 
The FRW metric is assumed to have been created behind a radially expanding shock--wave due to a great explosion
into a static, spherically symmetric universe modeled by a TOV metric.  Given these
assumptions, we have shown that conservation of energy at the shock then implies that the position $r$
 of the 
shock--wave 
is determined by equation 
(\ref{6.6}), where $r$ is the radial coordinate in the FRW universe behind the shock.  Equation
(\ref{6.6}) is coupled to equation (\ref{6.7}) for the TOV pressure $\bar{P},$ and the TOV energy density 
$\bar{Q}$
is then given by the formula (\ref{6.4c}).  In this section we assume that $w=R^4\bar{P}/\alpha$ 
lies on the critical orbit
$w=w_{crit}(S),$ ($S=\frac{R}{\alpha}$).  (This is justified by the fact that, according to 
Theorem \ref{theorem6.1},
this is the only orbit bounded for all
$R,$ and all orbits are asymptotic to this one as $R\rightarrow0.$)  By Theorem \ref{theorem6.1}, $w$ 
ranges between $\bar{\sigma}$ and $1/9$ along the critical orbit, and thus we have the apriori estimate

\begin{equation}
\label{8.1}
1/9\approx.1111<w<\bar{\sigma}\approx.1231.
\end{equation}
The only remaining piece of information missing is the initial condition for the shock--wave.  
At first one might think 
that this initial condition can be chosen arbitrarily, but as we have shown in the last section, 
the condition that the 
energy density be positive in front of the shock--wave, or that it be larger than the pressure 
in front of the shock,  
puts a constraint on the maximum shock position at a given time.  That is, assuming that 
$w$ lies on the critical
orbit implies that the hypotheses of Theorem
\ref{T7.3} hold, and thus condition (\ref{7.21}), 

\begin{equation}
\alpha r^2<\frac{(1-w)R^2}{3+\frac{\beta}{\alpha}R+w},\nonumber
\end{equation}
is equivalent to $\bar{Q}>0,$ and the condition (\ref{7.22}),

\begin{equation}
\alpha r^2<\left(\frac{1-\bar{\sigma}/3}{1+\bar{\sigma}}\right)\frac{(1-
w)R^2}{3+\frac{\beta}{\alpha}R+w},\nonumber
\end{equation}
is sufficient to guarantee that $\bar{Q}>\bar{P},$ at any given 
value of $R.$   Moreover, if
(\ref{7.21}) or (\ref{7.22}) hold at a given value $R=R_*,$ Theorem
\ref{T7.3} tells us that they continue to hold for all
$R>R_*.$  

Under the above assumptions, we now obtain estimates for the shock position.  To start, rewrite
(\ref{6.6}) as

\begin{eqnarray}
\frac{dr^2}{dR}&=&\frac{2(1-w)R}{\alpha\left(3+\frac{\beta}{\alpha}R\right)
\left(3+\frac{\beta}{\alpha}R+w\right)}\nonumber\\
&=&
\frac{2(1-w)\frac{\alpha}{\beta^2}R}{\left(R+\frac{\alpha}{\beta}(3+w)\right)\left(R+3\frac{\alpha}{\beta}\right)}.
\label{8.2}
\end{eqnarray}
Using (\ref{8.1}) in (\ref{8.2}) gives the estimate

\begin{eqnarray}
\frac{(1-w_+)\frac{2\alpha}{\beta^2}R}{\left(R+(3+w_+)\frac{\alpha}{\beta}\right)
\left(R+3\frac{\alpha}{\beta}\right)}
\leq\frac{dr^2}{dR}\leq
\frac{(1-w_-)\frac{2\alpha}{\beta^2}R}{\left(R+(3+w_-)\frac{\alpha}{\beta}\right)
\left(R+3\frac{\alpha}{\beta}\right)}
\nonumber\\
\label{8.3}
\end{eqnarray}
where $w_-=1/9\approx.1111<w_+=\bar{\sigma}\approx.1231.$  That is,

\begin{eqnarray}
\frac{\alpha}{\beta^2}
\frac{2(1-\bar{\sigma})R}{\left(R+(3+\bar{\sigma})\frac{\alpha}{\beta}\right)
\left(R+3\frac{\alpha}{\beta}\right)}
\leq\frac{dr^2}{dR}\leq
\frac{\alpha}{\beta^2}
\frac{(16/9)R}{\left(R+(\frac{28\alpha}{9\beta}\right)\left(R+3\frac{\alpha}{\beta}\right)}.
\nonumber\\
\label{8.3a}
\end{eqnarray}
Now by direct calculation, the solution to the ODE

\begin{eqnarray}
\label{8.4}
\frac{dr^2}{dR}=C\frac{R}{(R+A)(R+B)},
\end{eqnarray}
for positive constants $A,$ $B,$ and $C,$ is given by

\begin{eqnarray}
\label{8.5}
r^2=r_*^2+\ln\left[\left(\frac{R+A}{R_*+A}\right)^{\frac{AC}{A-B}}
\left(\frac{R+B}{R_*+B}\right)^{\frac{-BC}{A-B}}\right],
\end{eqnarray}
where inequalities can be substituted for 
equalities in (\ref{8.4}), (\ref{8.5}).   Applying this to (\ref{8.3}) gives the inequalities

\begin{eqnarray}
\label{8.6}
r^2-r_*^2\geq \ln\left[\left(\frac{R+A_+}
{R_*+A_+}\right)^{a_+}
\left(\frac{R+B_+}{R_*+B_+}\right)^{b_+}\right],
\end{eqnarray}

\begin{eqnarray}
\label{8.7}
r^2-r_*^2\leq \ln\left[\left(\frac{R+A_-}
{R_*+A_-}\right)^{a_-}
\left(\frac{R+B_-}{R_*+B_-}\right)^{b_-}\right],
\end{eqnarray}
where

\begin{eqnarray}
A&=&(3+w)\frac{\alpha}{\beta},\nonumber\\
B&=&3\frac{\alpha}{\beta},\nonumber\\
a&=&\frac{2(3+w)(1-w)}{w}\frac{\alpha}{\beta^2},\nonumber\\
b&=&-\frac{6(1-w)}{w}\frac{\alpha}{\beta^2},\nonumber\\
\end{eqnarray}
and $A_-,$ $A_+,$ are obtained by substituting $w_-,$ $w_+$ for $w,$ respectively, in the above expressions, etc.

We now evaluate $\alpha$ and $\beta$ in terms of the present value of the Hubble constant $H_0$ and the 
observed microwave background radiation temperature $T_0.$  Here we let subscript zero denote value at
present time in the FRW metric,
and WLOG we assume that $R_0=1.$   Recall that 
the FRW equation (\ref{2.3b}) for $k=0$
can be written as

$$
H^2=\left(\frac{\dot{R}}{R}\right)^2=QR^2,
$$
so that

$$
H_0=\sqrt{Q_0},
$$
where $Q_0$ denotes the present value of the (scaled) energy density in the universe at present time.
By (\ref{5.12}), 

$$
Q_0=3\alpha+\beta,
$$
where $3\alpha$ is the energy density of radiation at present time, and $\beta$ is the energy density of matter at
present time.  Let $T\equiv T(R)$ denote the temperature of 
radiation.  Then (\ref{5.16}) is 

\begin{equation}
T=\left(\frac{3\alpha}{\hat{a}}\right)^{1/4}\frac{1}{R},\nonumber
\end{equation}
where (\ref{SB3}) gives

$$
\hat{a}=\frac{8\pi{\cal G}a}{3c^4}.
$$
Setting $R_0=1$ and solving (\ref{5.16}) for $\alpha$ gives

\begin{equation}
\label{8.9}
\alpha=\frac{\hat{a}}{3}T_0^4,
\end{equation}
and using this in (\ref{5.12}) gives

\begin{equation}
\label{8.10}
\beta=Q_0-3\alpha=H_0^2-\hat{a}T_0^4.
\end{equation}
We evaluate the above constants using the values, (\cite{wein}),

\begin{eqnarray}
&\frac{\cal G}{c^2}&=7.425\times 10^{-29}\ cm\ g^{-1},\label{8.11}\\
&c&=2.997925\times 10^{10}\ cm\ sec^{-1},\label{8.12}\\
&lty&=9.4605\times 10^{17}\ cm,\label{8.13}\\
&mpc&=10^6\ pc=3.2615\times 10^{6}\ lty.\label{8.14}\\
&a&=7.5641\times 10^{-15}\ erg\ cm^{-3}\ K^{-4},\label{8.15}\\
&H_0&=100h_0\ km\ sec^{-1}\ mpc^{-1},\label{8.16}\\
&T_0&=2.736\ ^oK,\label{8.16a}
\end{eqnarray}
Here, ${\cal G}$ is Newton's gravitational constant, $c$ the speed of light, $lty$ is lightyear, $mpc$
is megaparcec, $^oK$ is degrees Kelvin, $a$ is the Stefan-Boltzmann
constant, $T_0$ is the observed microwave background radiation temperature \cite{peeb},
and $H_0$ is Hubble's constant, where $h_0$ is generally accepted to be between $.5$ and unity.
(We take $h_0\approx.55$ as a recently quoted value.)  Using these values we calculate

\begin{equation}
\label{8.17}
\hat{a}=4.6852\times 10^{-27}\ lty^{-2}\ K^{-4},
\end{equation}

\begin{equation}
\label{8.18}
H_0=1.023h_0\times 10^{-10}\ lty^{-1}.
\end{equation}

Using the above values we obtain from (\ref{8.9}) and (\ref{8.10}) that

\begin{equation}
\label{8.21}
\frac{\alpha}{\beta}=\frac{1/3}{\left(\frac{H_0^2}{\hat{a}T_0^4}\right)+1}
\approx\frac{\hat{a}}{3H_0^2}T_0^4=\frac{1.492T_0^4}{h_0^2}\times 10^{-7},
\end{equation}
and 

\begin{eqnarray}
\label{8.22}
\frac{\alpha}{\beta^2}&=&\frac{\hat{a}T_0^4}
{3\left[\hat{a}T_0^4\left(\frac{H_0^2}{\hat{a}T_0^4}\right)-1\right]}
\approx\left(\frac{\hat{a}}{3H^2_0}\right)\frac{T_0^4}{H_0^2}\\
&=&(1.492\times 10^{-7})\frac{T_0^4}{h_0^2H_0^2}
=\frac{8.34\times 10^{-6}}{h_0^2H_0^2},
\end{eqnarray}
where we used $\frac{\hat{a}T_0^4}{H^2_0}<<1$ at the approximate equality.
Using these values we can evaluate:

\begin{eqnarray}
A_+&=&(3+.1231)\frac{\alpha}{\beta}=(4.66\times10^{-7})\frac{T_0^4}{h_0^2},
\label{8.23}\\
B_+&=&3\frac{\alpha}{\beta}=(4.48\times10^{-7})\frac{T_0^4}{h_0^2},\nonumber\\
a_+&=&\frac{2(3+.1231)(1-.1231)}{.1231}\frac{\alpha}{\beta^2}=
(6.639\times10^{-6})\frac{T_0^4}{h_0^2H_0^2}.
\nonumber\\
b_+&=&-\frac{6(1-.1231)}{.1231}\frac{\alpha}{\beta^2}=
(6.377\times10^{-6})\frac{T_0^4}{h_0^2H_0^2},\nonumber
\end{eqnarray}

\begin{eqnarray}
A_-&=&(3+1/9)\frac{\alpha}{\beta}=
(4.64\times10^{-7})\frac{T_0^4}{h_0^2},\label{8.24}\\
B_-&=&3\frac{\alpha}{\beta}=(4.48\times10^{-7})\frac{T_0^4}{h_0^2},\nonumber\\
a_-&=&\frac{2(3+1/9)(1-1/9)}{1/9}\frac{\alpha}{\beta^2}=
(7.427\times10^{-6})\frac{T_0^4}{h_0^2H_0^2},\nonumber\\
b_-&=&-\frac{6(1-1/9)}{1/9}\frac{\alpha}{\beta^2}=
(7.162\times10^{-6})\frac{T_0^4}{h_0^2H_0^2}.\nonumber
\end{eqnarray}

Now, assuming that the uncoupling of matter and radiation occured at a temperature less than
$4000$ degrees Kelvin, \cite{wein}, it follows from (\ref{5.16}) that

\begin{equation}
\label{8.25}
R_*\geq 2.2/4000=6.75\times 10^{-4},
\end{equation}
and so it follows that we can essentially neglect the $A$'s and $B$'s in estimates (\ref{8.6}) and
(\ref{8.7}), given their small values in (\ref{8.23}) and (\ref{8.24}), and assuming this, 
estimates (\ref{8.6}) and (\ref{8.7}) reduce to,
\begin{equation}
\label{8.26}
(a_++b_+)\ln(1/R_*)\leq r^2-r_*^2\leq (a_-+b_-)\ln(1/R_*).
\end{equation}
Using (\ref{8.23}) and (\ref{8.24}) to estimate (\ref{8.26}) gives the following 
estimate for the distance the shock--wave must have traveled between $r=R_*$ and $R=1$ as
predicted by our model:

\begin{equation}
\label{8.27a}
\frac{(2.62\times 10^{-7})T_0^4}{h_0^2H_0^2}
\ln\left(\frac{1}{R_*}\right)
\leq r^2-r_*^2\leq
\frac{(2.65\times 10^{-7})T_0^4}{h_0^2H_0^2}
\ln\left(\frac{1}{R_*}\right).
\end{equation}
Here the distance $r$ is given in terms of the Hubble length

\begin{equation}
\label{8.28}
H_0^{-1}\approx \frac{.98}{h_0}\ \times10^{10}.
\end{equation}
In particular, (\ref{8.27a}) shows that, in this shock--wave model, the quantity $r^2-r_*^2$ is essentially independent of 
the starting position $r_*.$  

As an example, if we take
$h_0=.55,$ $T_0=2.736^oK$ and $R_*=2.7/4000$ in (\ref{8.27a}), we obtain the estimate

\begin{equation}
\label{8.29}
r^2-r_*^2\approx\left(\frac{.019}{H_0}\right)^2.
\end{equation}
In the standard interpretation of the FRW metric in Cosmology, 
the galaxies are in freefall, and
traverse geodesics $r=Const.$  Thus we can interpret $r^2-r_*^2$ in 
(\ref{8.27a}) as the (squared) distance that
the shock--wave travels over and above the motion due to freefall, a result of the fact that mass and momentum
are driven across the shock--wave as it evolves outward. We conclude that the distance the shock--wave has 
traveled, 
(over and above freefall),
between $R=R_*=2.7/4000$ and 
$R=1,$  as predicted by this model,
is approximately $.019$ of the Hubble length. 
(Recall that $\bar{r}=R(t)r$ measures distance in
lightyears for the three dimensional space at fixed time 
$t$ in the $k=0$ FRW metric.) 

We now discuss the initial condition $r=r_*$ at $R=R_*.$   We saw in 
(\ref{7.22})
that the condition $\bar{Q}>\bar{P}$ put constraints on the maximal 
shock position at each value of
$R.$  Using the value $\bar{\sigma}=.1231$ in (\ref{7.22}) gives the 
inequality

\begin{equation}
\label{8.30}
r_*^2<\frac{.759R^2}{(3.11)\alpha+\beta R}=\frac{.759}
{1+\left[\frac{(4.64\times10^{-7})T_0^4}{h_0^2R_*^2}\right]}\frac{R_*}{H_0^2}.
\end{equation} 
Estimate (\ref{8.30}) is the bound on the initial shock 
position, imposed by 
$\bar{Q}>\bar{P},$ in terms of the Hubble length.
Putting (\ref{8.30}) together with (\ref{8.27a}), we conclude that 
the maximal distance $r_{max}$ from the 
shock--wave
to the center of the explosion $r=0$ {\em at present time $R=1$}, 
given as a function of starting time $R_*,$ $2.7/4000\leq R_*1$, 
(assuming the 
shock--wave started at 
position $r=r_*$ at $R_*\geq2.7/4000,$ and such that $r_*$ is restricted 
by (\ref{8.30}) 
so that $\bar{Q}>\bar{P}$ for all $R>R_*$), 
is predicted by this model to be

\begin{equation}
\label{8.33}
r_{max}\approx H_0^{-1}\sqrt{\frac{.76}
{1+\left[\frac{(4.6\times10^{-7})T_0^4}{h_0^2R_*^2}\right]}R_*+
(2.6\times10^{-7})\frac{T_0^4}{h_0^2}
\ln{\left(\frac{1}{R_*}\right)}}.
\end{equation}
For example, taking the value $h_0=.55$ and $T_0=2.736^oK$ gives the formula

\begin{equation}
\label{8.34}
r_{max}\approx H_0^{-1}\sqrt{\frac{.76}
{1+\left[\frac{(8.5\times10^{-5})}{R_*^2}\right]}R_*+
(4.9\times10^{-5})
\ln{\left(\frac{1}{R_*}\right)}}.
\end{equation}
This function is plotted in Figure 2.  Putting (\ref{8.33}) together with
(\ref{8.27a}) we obtain the following upper and lower bounds for the 
shock position $r$ at present time $R=1$ assuming that it
starts at $R=R_*,$ and such that $\bar{Q}>\bar{P}$ holds 
for all $R\geq R_*:$

\begin{eqnarray}
r&\geq&H_0^{-1}\left\{(5.1\times10^{-4})\frac{T_0^2}{h_0}
\sqrt{\ln{\left(\frac{1}{R_*}\right)}}\right\},\label{final1}\\
r&\leq& 
H_0^{-1}\sqrt{\frac{.76}
{1+\left\{\frac{(4.6\times10^{-7})T_0^4}{h_0^2R_*^2}\right\}}R_*+
(2.6\times10^{-7})\frac{T_0^4}{h_0^2}
\ln{\left(\frac{1}{R_*}\right)}}.\nonumber\\
\label{final2}
\end{eqnarray}

\section{The Case of Pure Radiation, $\beta=0$}
\label{Sect8} 
\setcounter{equation}{0}

As a point of comparison, in this section we redo the calculation of the 
shock position under the assumption $\beta=0$ in (\ref{5.12}); that is, 
under the assumption 
that the energy density $Q$ is due entirely to radiation.  (See 
\cite{smolte2}
and the solution discussed at the end Section \ref{Sect3}.)  Thus assume
that $\alpha=\hat{a}T_0^4/3$ is as given in (\ref{8.9}), but that $\beta=0.$
We estimate the position of the shock--wave in this model at the time
$R=1,$ where $T=T_0.$  Now of course, since $\beta$ is determined in (\ref{8.10})
from $H_0$ in the above analysis, the value of $\frac{\dot{R}}{R}$ in the
pure radiation model will not coincide with $H_0$ at the time when
$T=T_0.$  Nevertheless, for 
comparison purposes,  we shall estimate the radial position of the shock--wave
in the pure radiation model at time $R=1$  
in terms of the Hubble length $H_0^{-1}$ given in (\ref{8.28}).  

In the case $\beta=0,$  the 
constraint (\ref{7.22}) that guarantees $\bar{Q}>\bar{P}$ reduces to

\begin{equation}
\label{9.1}
\alpha r_*^2<\left(\frac{(1-\bar{\sigma}/3)(1-w)}{(1+\bar{\sigma})(3+w)}\right)
R_*^2,
\end{equation}
and the critical orbit becomes $w\equiv\bar{\sigma}.$  Using 
$w=\bar{\sigma}\approx.1231$ in (\ref{9.1}) gives

\begin{equation}
\label{9.2}
r_*<\frac{.49}{\sqrt{\alpha}}R_*.
\end{equation}
(Note that in the alternative case $\alpha=0,$ the case of 
pure matter, 
the RHS of (\ref{7.22}) tends to infinity, and thus (\ref{7.22}) places no
constraint on the shock position.  This is consistent with the fact that
when $\alpha=0,$ the pressure is zero, and the shock--wave reduces to 
a contact discontinuity.  For example, $\bar{Q}=0,$ $\bar{P}=0,$ solves 
the shock equations (\ref{4.39}), (\ref{4.40}) and 
it is not difficult to 
show that the solution of the shock
equations in this case reduces to the $k=0$ version of the Oppenheimer--Snyder model,
first presented in \cite{smolte2}.  In these
Oppenheimer--Snyder models, there are no constraints on the shock position
corresponding to (\ref{7.22}).)   

Setting $\beta=0$ and $w=\bar{\sigma}$
in (\ref{6.6}) gives

\begin{equation}
\label{9.3}
\frac{dr^2}{dR}=\frac{2(1-\bar{\sigma})}{3(3+\bar{\sigma})}\frac{R}{\alpha}
\end{equation}
as the differential equation for the shock position.  Integrating gives

\begin{equation}
\label{9.4}
r^2=\frac{(1-\bar{\sigma})}{3(3+\bar{\sigma})}\frac{R^2}{\alpha}+r_*^2.
\end{equation} 
Using (\ref{9.2}) for the maximum value of $r_*$ yields the following bounds 
on the shock position $r$ at the time $R=1$ when $T=T_0$ that are 
analagous to 
(\ref{final1}) and (\ref{final2}) and apply when $\beta=0:$

\begin{equation}
\label{9.5}
\frac{1}{\sqrt{\alpha}}\sqrt{\frac{(1-\bar{\sigma})}{3(3+\bar{\sigma})}}
\leq r\leq\frac{1}{\sqrt{\alpha}}\sqrt{\frac{(1-\bar{\sigma})}{3(3+\bar{\sigma})}
+.24R_*}.
\end{equation}
>From (\ref{8.9}) and (\ref{6.21}) it follows that
\begin{equation}
\label{9.6a}
\frac{1}{\sqrt{\alpha}}=118h_0H_0^{-1}.
\end{equation}
Using this value together with the value 
$\bar{\sigma}=.1231$ in (\ref{9.5}) yields

\begin{equation}
\label{9.7}
\frac{36h_0}{H_0}\leq r\leq\frac{36h_0\sqrt{1+2.5R_*}}{H_0}.
\end{equation}
Note that the shock position at $R=1$ that applies to the exact solution
given in \cite{smolte2}, which was discussed in detail at the end of
Section \ref{Sect3}, is the case $R_*=0$ in (\ref{9.7}).  

\begin{figure}\label{Fig1}
\psfig{file=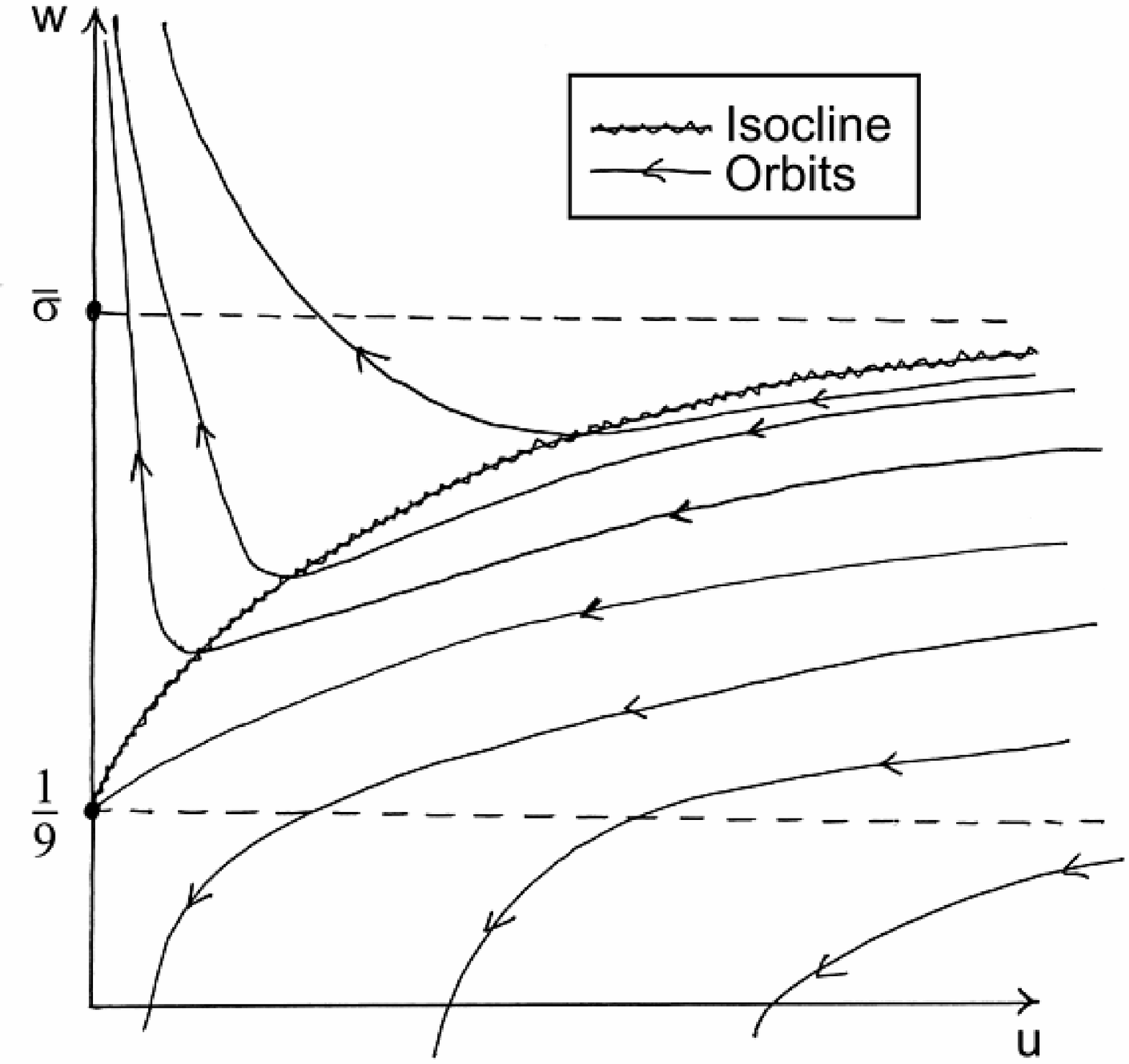,height=5.0in,width=5.0in}
\caption{}
\end{figure}

\begin{figure}\label{Fig2}
\psfig{file=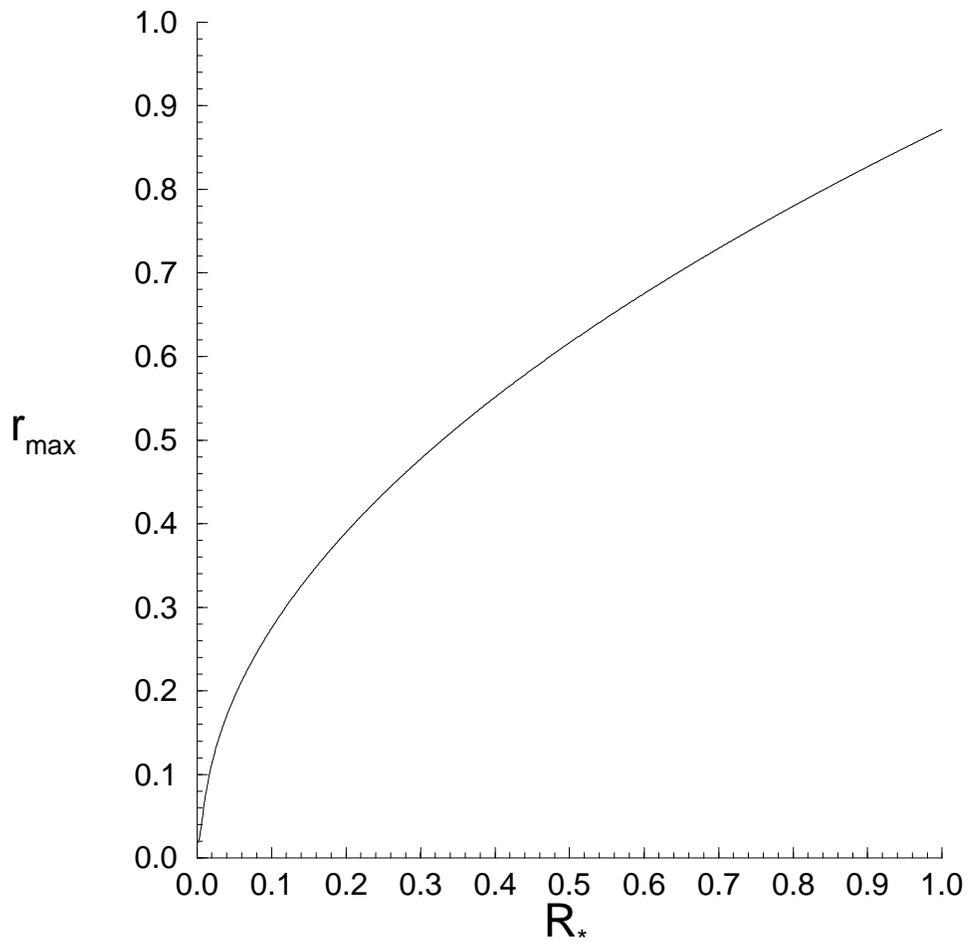,height=5.0in,width=5.0in}
\caption{$r_{max}$ is in units of $H_0^{-1},$ $H_0=100h_0\frac{km}{sec~mpc},$ $h_0=.55$}
\end{figure}

\end{document}